\documentclass[referee]{raa}            

\usepackage{graphicx,times}             
\usepackage{natbib}
\usepackage{amssymb,amsmath}
\bibpunct{(}{)}{;}{a}{}{,}

\usepackage[pagebackref=true]{hyperref}

\begin{document}

  \title{Atmospheric parameters and kinematic information for the M giants stars from LAMOST DR9
}

   \volnopage{Vol.0 (20xx) No.2022-0433}      
   \setcounter{page}{1}          

   \author{Dan Qiu 
      \inst{1,2}
   \and Hao Tian 
      \inst{3,1}
   \and Jing Li
      \inst{4}
   \and Chao Liu
      \inst{1,2}
   \and Lin Long
      \inst{4}
   \and Jian-Rong Shi
      \inst{5,2}
   \and Ming Yang
      \inst{1}
   \and Bo Zhang
      \inst{1}
   }

   \institute{Key Laboratory of Space Astronomy and Technology, 
National Astronomical Observatories, 
Chinese Academy of Sciences, 
Beijing 100101, PR China; {\it tianhao@nao.cas.cn}\\
        \and
             University of Chinese Academy of Sciences, 100049, PR China\\
        \and 
             Institute for Frontiers in Astronomy and Astrophysics, Beijing Normal University, Beijing, China
        \and
             School of Physics and Astronomy, China West Normal University,
             1 ShiDa Road, Nanchong 637002, PR of China\\
        \and 
             Key Laboratory of Optical Astronomy, National Astronomical Observatories,
             Chinese Academy of Sciences, Beijing 100101, PR China;
\vs\no
   {\small Received 2022 November 17; accepted 2023 February 7}}

\abstract{ A catalog of more than 43,000 M giant stars has been selected by 
Li et al. from the ninth data release of LAMOST.
Using the data-driven method SLAM, we obtain the  stellar parameters 
($T_{\mathrm{eff}}$, $\mathrm{log}\,g$, $[\mathrm{M/H}]$, $[\alpha/\mathrm{M}]$) 
for all the M giant stars with uncertainties of $57$ K, $0.25$ dex, 
$0.16$ dex and $0.06$ dex at SNR$>$100, respectively. With those 
stellar parameters, we constrain the absolute magnitude in $K-$band, 
which brings distance with relative uncertainties 
around $25\%$ statistically. Radial velocities 
are also calculated by applying cross correlation on the spectra between
$8000$ \AA  \,and $8950$ \AA \, with synthetic 
spectra from ATLAS9, which covers the Ca II triplet. 
Comparison between our radial velocities and those from APOGEE DR17 and Gaia
DR3 shows that our radial velocities
have a system offset and dispersion around $1$ and $4.6$ km s$^{-1}$,
respectively. With the distances and radial velocities combining with the 
astrometric data from Gaia DR3, we calculate the full 6D 
position and velocity information, which are able to be 
used for further chemo-dynamic studies on the disk and substructures 
in the halo, especially the Sagittarius Stream.
\keywords{methods: statistical -- stars: evolution, fundamental parameters -- Galaxy: stellar content}
}

   \authorrunning{D. Qiu et al }            
   \titlerunning{Chemical and Kinematic information for the M giant sample}  

   \maketitle

%
%
\section{Introduction}           
\label{sect:intro}
M giant stars are the kind of stars with high luminosity and low temperature, such as
the tip of the red giant branch stars (tRGB stars), $C$-, $O$- and extreme asymptotic 
giant branch stars (AGB stars) and red super giant stars.  
On the first hand, the brightness means that 
they are able to be used to trace the distant volumes, 
which makes them a good tracer to reveal the accretion and merger events in the Milky Way 
by discovering and identifying the remnants of the relatively metal rich stellar streams in the halo,
especially for the Sagittarius system \citep{Ibata1994Natur.370..194I}, which is still 
suffering tidal 
disruption \citep{Newberg-2002, Belokurov-2014, Koposov-2015, Li-2016a, Li-2016b}.
\cite{Majewski-2003} selected the M  giant stars from 2MASS. Those samples 
clearly represented the Galactic disk and satellite galaxies, such as
the Magellanic Clouds and the Sagittarius dwarf spheroidal galaxy. 
\cite{Li-2016a} also used the M giant stars
to map the Sagittarius Stream and 
revealed more distant structure. 
On the other hand,  the low
temperatures indicate that most of the flux are distributed at the long wavelength bands, such 
as the $K$-band in 2MASS system.
Therefore, a further advantage is that the M giant stars suffer less extinction. This provides 
the opportunity to study the outer volumes of the disk with low latitude.

Though the M giant stars have significant advantages, they are not widely used for the halo 
and disk studies, especially comparing with the K giant stars 
\citep{LiuC2017RAA....17...96L,Xu2018MNRAS.473.1244X,TianH2019ApJ...871..184T, TianH2020ApJ...899..110T,Xu2020ApJ...905....6X}. The first reason is that there are much 
fewer M giant stars than the K giant stars. The other one  is that,
because of the low temperature, there are molecular absorption bands in their
spectra, which bring difficulties to constrain the stellar physical parameters, such 
as the abundances and the radial velocities
which are quite important 
for further studies on the structures in the Milky Way.
In recent years, with the development of observation equipment, 
many large surveys like Large Sky Area Muliti-Object Fiber 
Spectroscopic Telescope \citep[LAMOST;][]{Wang-1996,Cui-2012, Deng-2012, Zhao-2012, Luo-2012, Su-2004, Yan-2022}
Sloan Digital Sky Survey \citep[SDSS;][]{Ahumada-2020}, collected a 
large amount of photometric and low resolution spectral data of 
M-type stars. Meanwhile kinds of stellar parameter pipelines have been developed.
However, the Stellar Parameter Pipelines, 
like the LAMOST Stellar Parameter Pipelines \citep[LAPS;][]{Wu-2014,Luo-2015}, 
which designed based on the University of Lyon Spectroscopic Analysis Software
\citep[ULYSS;][]{Koleva-2009}, is unable to derive accurate stellar
parameters for these M giant with low resolution spectra because of those molecular
bands. Special attention should be paid on those low temperature
stars, such as the pipeline for the Apache Point Observatory Galactic Evolution 
Experiment \citep[APOGEE DR17;][]{Abdurro-2022}.

The most common and efficient method to determine the stellar parameters 
is to fit the observed spectra with the 
synthetic spectra, which has been successfully applied on the RGB stars for
decades.
\citet{Bizyaev-2006,Bizyaev-2010} determined the stellar parameters of hundreds of red giants with 
high resolution spectra  by comparing the observed spectra to a synthetic stellar spectra library 
ATLAS9 \citep{Kurucz-1993}, including few low temperature stars.
With similar method,  \citet{Carlin-2018} determined the ${T_{\mathrm{eff}}}$, 
[Fe/H] and log\,$g$ of 42 K/M giants with high resolution 
spectral (R$\sim$ 67500) from Gemini Remote Access to CFHT ESPaDOnS Spectrograph
\citep[GRACES;][]{Tollestrup-2012,Chene-2014}, 
using a synthetic spectra atmospheric models generated 
with \citet{Castelli-2003} and Dartmouth isochrones \citep{Dotter-2008}. 
The typical uncertainties on ${T_{\mathrm{eff}}}$, [Fe/H] and log\,$g$ are 115 \,K, 0.09\,dex 
and 0.18\,dex, respectively.
More recently, \citet{Ding-2022} derived stellar atmospheric parameters of LAMOST M-type stars from 
MILES library interpolator by applying the $\chi^2$ minimization performed by the ULySS package. For M 
giants, the uncertainties of ${T_{\mathrm{eff}}}$, log$g$ and [Fe/H] are 58 K, 0.19 dex and 0.26 dex, respectively.

With rapid development, the machine learning has been applied on deriving the stellar parameters 
frequently in recent years \citep{Howard-2017, Antoniadis-2020, Galgano-2020,Zhang-2020}.  More recently,
\citep{Wang-2020} designed a neural network model, 
named SPCANet, to determined the ${T_{\mathrm{eff}}}$, log\,$g$ and 13 chemical abundances for medium
resolution spectroscopy from LAMOST Medium Resolution Survey (MRS) data
sets (R $\sim$7500) \citep{LiuChao2020arXiv200507210L},
 including many M giant spectra. The precision of 
${T_{\mathrm{eff}}}$, log\,$g$ and [Fe/H] are 119 K, 0.17 dex 
and 0.06 dex, respectively.

In this work, we use a data-driven method Stellar LAbel Machine (SLAM),
which is developed by \citet{Zhang-2020} to derive the stellar parameters of M giants
from low resolution (R $\sim$ 1800) spectra of LAMOST,
including  ${T_{\mathrm{eff}}}$, [M/H] and log\,$g$. SLAM 
has showed good performance in deriving stellar parameters.
e.g.,\citet{Zhang-2020} used SLAM to determined $T_{\mathrm{eff}}$, log$\,g$ and [Fe/H] from 
low-resolution spectra for $\sim$ 1 million LAMOST DR5 K giants  with random uncertainties are 50 K, 0.09 dex and 0.07 dex, respectively. 
\citet{Li-2021} measured  $T_{\mathrm{eff}}$ and [M/H] of M dwarfs by training SLAM 
with LAMOST low-resolution spectra and APOGEE stellar labels, the  $T_{\mathrm{eff}}$ 
and [M/H] are in agreement to within 50 K and 0.12 dex compare to the 
APOGEE observation.
\citet{Guo-2021} adopted SLAM to predict  $T_{\mathrm{eff}}$, log$\,g$, [M/H] and 
projected rotational velocity ($v\sin i$) for 3931 early-type stars from LAMOST low-resolution survey. The uncertainties of  $T_{\mathrm{eff}}$, log$\,g$ and $v\sin i$ are 1642 K, 0.25 dex 
and $42\,\mathrm{km}\,{{\rm{s}}}^{-1}$, respectively. They also determined the above four parameters 
by using SLAM for 578 early-type stars from LAMOST medium-resolution 
survey (MRS). The uncertainties are 2185 K, 0.29 dex and  $11\,\mathrm{km}\,{{\rm{s}}}^{-1}$ for $T_{\mathrm{eff}}$, log$\,g$ and $v\sin i$ , respectively.

This paper is organized as follows: a brief description about the sample will be presented in
the Section \ref{sect:Data}. In Section \ref{sect:Method} we will show the results of the radial velocities and the stellar parameters.
The validation of the parameters are also discussed in the Section \ref{sect:Method}. Then the 
application of this value added catalog will be showed in the Section \ref{sect:App}. Finally, the summary 
will be given in the Section \ref{sect:conclu}.

\section{Data}\label{sect:Data}

\subsection{M giants}\label{sect:M_gaints}

Millions of spectra have been obtained by LAMOST in the last 10 years 
\citep{Cui-2012, Deng-2012,Zhao-2012}, including thousands of M giant stars. 
Though there are many molecular absorption bands in the spectra, \cite{Zhong2015RAA....15.1154Z}
and \cite{LiJ-2019} have successfully selected the M giant stars using similar method on 
the spectra from LAMOST.
Recently, using the similar method, Li et al. (in preparation) have selected more than 43,000 reliable 
M giant stars from LAMOST DR9. Those M giant stars are separated from the M dwarf stars using the spectra
index of TiO5 versus that of CaH2+CaH3, which has been proved to be a quite efficient way
\citep{Zhong2015RAA....15.1154Z}.
The M type giant and dwarf stars are behaved at two different
clumps in the color-color diagram, $(W1-W2)_0$ versus $(J-K)_0$.
At last the contamination of few white dwarf stars and those dwarf stars located in the overlapped region 
with the M giant stars in the spectra index diagram and the color-color diagram
can be further reduced by applying the distance provided by Gaia DR3 \citep{Zhong2019ApJS..244....8Z}.
After all those selections, more than 43000 M giant stars are left, which will be 
used in this manuscript.

\begin{figure}
\centering
\includegraphics[width=0.3\textwidth,trim=0.cm 6cm 2.0cm 1.cm, clip]{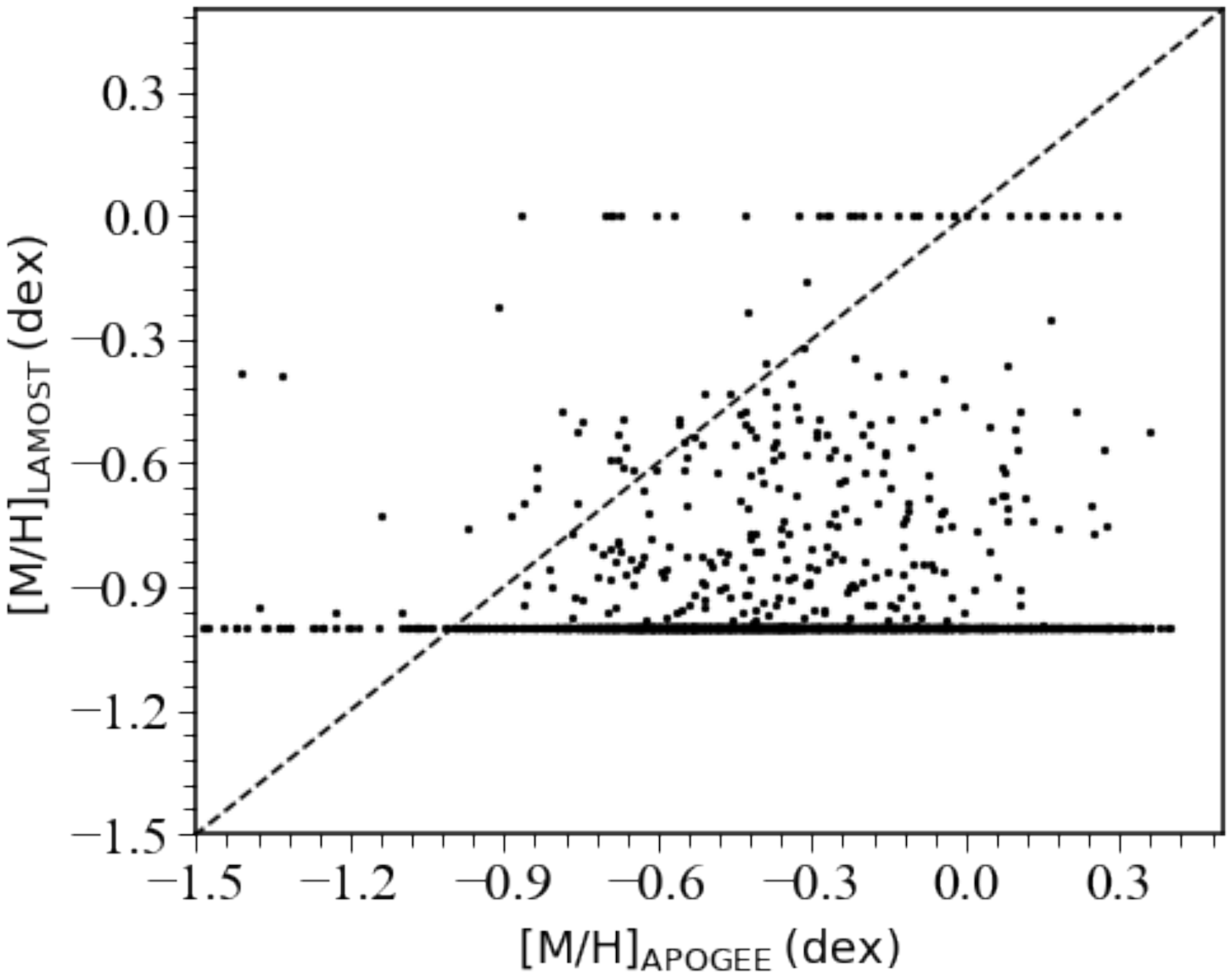}
\includegraphics[width=0.3\textwidth,trim=0.cm 6cm 2.0cm 1.cm, clip]{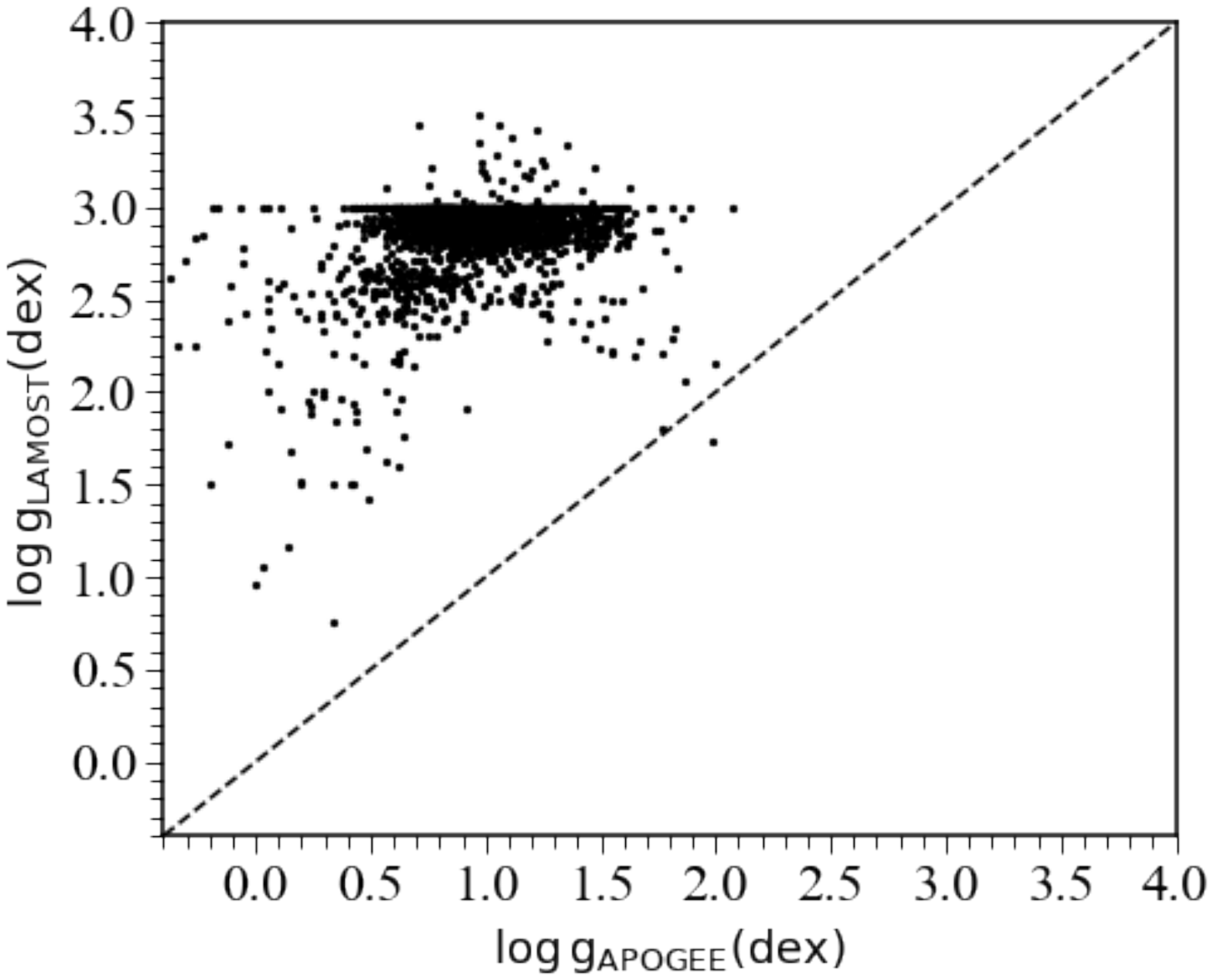}
\includegraphics[width=0.3\textwidth,trim=0.cm 6cm 2.0cm 1.cm, clip]{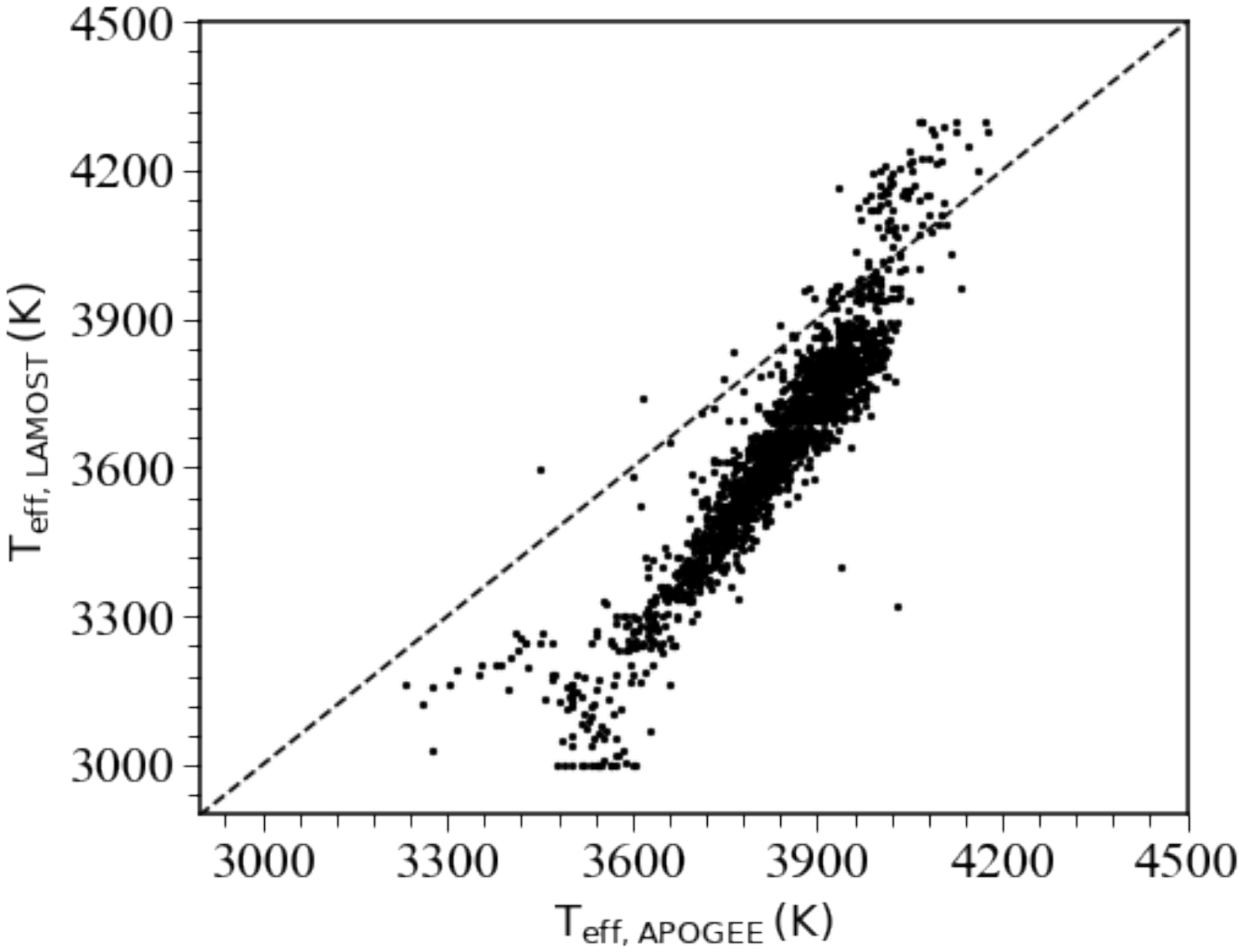}
\caption{The comparison of metallicity, surface gravity and effective temperature between LAMOST and APOGEE of 2123 M giant stars, respectively.
}\label{fig:Teff_X_err}
\end{figure}

There are 28610 M giant stars both in Li et al (in preparation) and LAMOST officially released M giant stars. 
We cross-match these 28610 M giants with APOGEE DR17, and obtain 2123 common stars
with both signal-to-noise ratios (SNR) from LAMOST and 
APOGEE  larger than 50. 
The SNR from LAMOST is defined as $\mathrm{SNR}=(\sum_{i=0}^{N} flux_i*\sqrt{invvar_i})/N$, where
$flux_i$ and $invvar_i$ are the flux and inverse 
variance of $i$th pixel of a spectrum, $N$ is the number of 
pixels of the corresponding spectrum. 
In Figure \ref{fig:Teff_X_err},
we represent the comparison of metallicity, surface gravity and 
temperature between LAMOST and APOGEE of those 2123 stars, respectively.
It indicates that there are great 
differences in these three stellar parameters between 
LAMOST and APOGEE, especially for [M/H] and 
log$\,{\mathrm{g}}$. 1840 of those common stars are not provided available parameters, 
but the constants -1 or 0 for the metallicity. That means the LAMOST pipeline cannot 
give reliable metallicity and surface gravity for those low temperature stars.
Due to the low temperature, the parameters from APOGEE with near-infrared spectra are more reliable.
Therefore, the results from APOGEE are adopted during our constraint on the stellar parameters.

\subsection{Training set}\label{sect:Train_set}
In this work, we train the SLAM model to predict the stellar parameters of M giant stars. The training set is used to train the model of the stellar labels versus the spectra. This requires that
the spectra of the training set should have high signal to noise ratios (SNRs) 
and  the accurate labels to each 
spectrum, e. g. [M/H], [$\alpha$/M], ${T_{\mathrm{eff}}}$ and log$\,{\mathrm{g}}$ in this work. To this way, we cross-match
the whole sample of M giant stars with APOGEE DR17  \citep{Abdurro-2022} and obtain a 
common catalog of 4473 M giant stars. 
With following criteria, we further constrain the accuracy of
the stellar parameters   ${T_{\mathrm{eff}}}$, log\,$g$,  [M/H] and [$\alpha$/M] , at last
  3670  M giant stars are left for the training set, including the accurate 
stellar parameters from APOGEE DR17 and high SNR spectra from LAMOST DR9.
\begin{enumerate}
\item[1.] $\mathrm{SNR}>50$
\item[2.] $\sigma_{[\mathrm{M/H}]}<0.1$
\item[3.] $\sigma_{T_{\mathrm{eff}}}<20$
\item[4.] $\sigma_{\mathrm{log}\,g}<0.1$
\item[5.] $\sigma_{[\mathrm{\alpha/M}]}<0.05$
\end{enumerate}
where SNR is the mean signal-to-noise ratio of LAMOST spectra as described in subsection \ref{sect:M_gaints},
$\sigma_{[\mathrm{M/H}]}$, $\sigma_{T_{\mathrm{eff}}}$, $\sigma_{\mathrm{log}\,g}$ and $\sigma_{[\mathrm{\alpha/M}]}$ are the uncertainties of the metallicity [M/H], the effective temperature $T_{\mathrm{eff}}$, the surface gravity $\mathrm{log}\,g$ and the alpha abundance ${[\mathrm{\alpha/M}]}$ of the 
M giant stars, which are provided by APOGEE DR17. 

\begin{figure}
\centering
\includegraphics[width=0.45\textwidth,trim=0.cm 0cm 2.5cm 1.cm, clip]{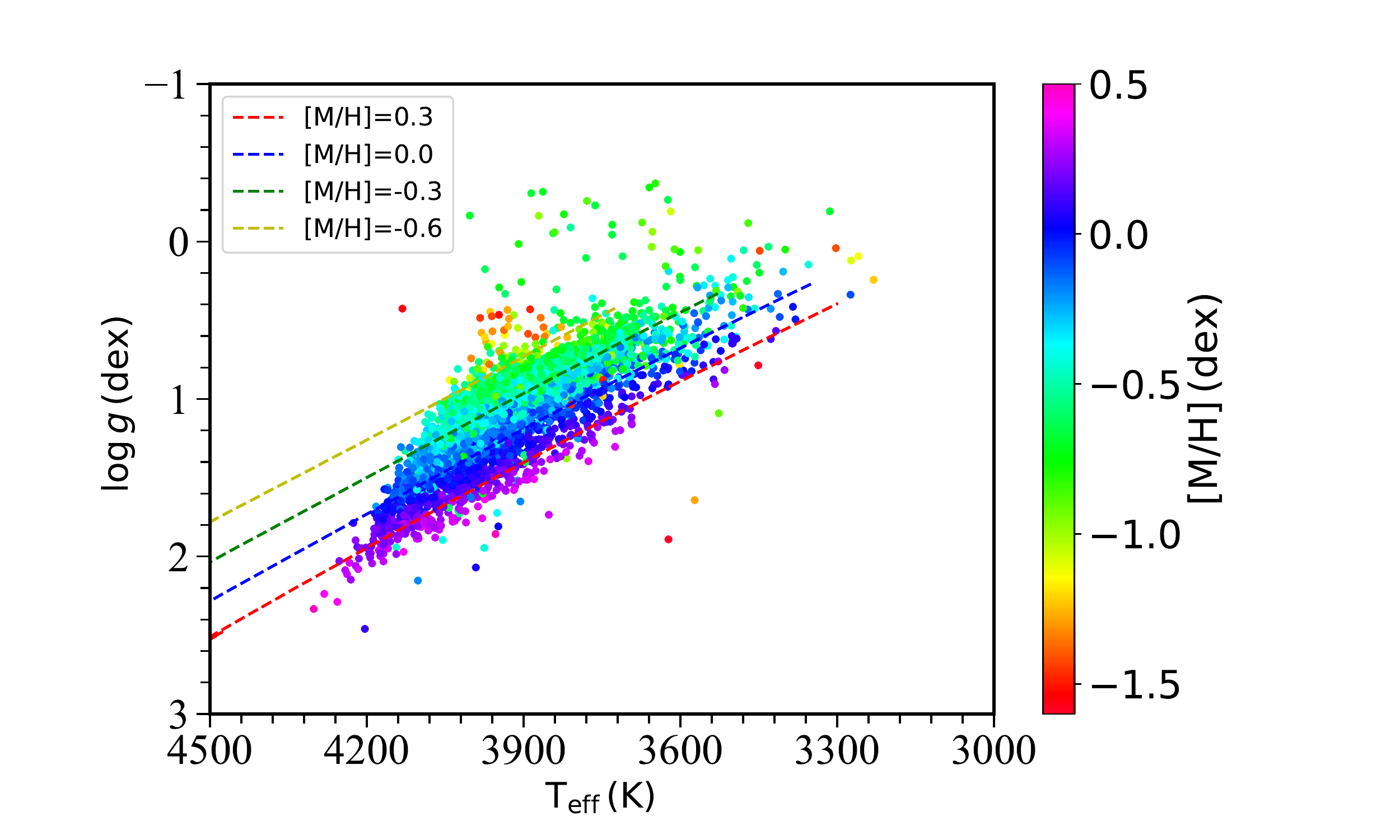}
\includegraphics[width=0.45\textwidth,trim=3.5cm 2.2cm 2.8cm 1.cm, clip]{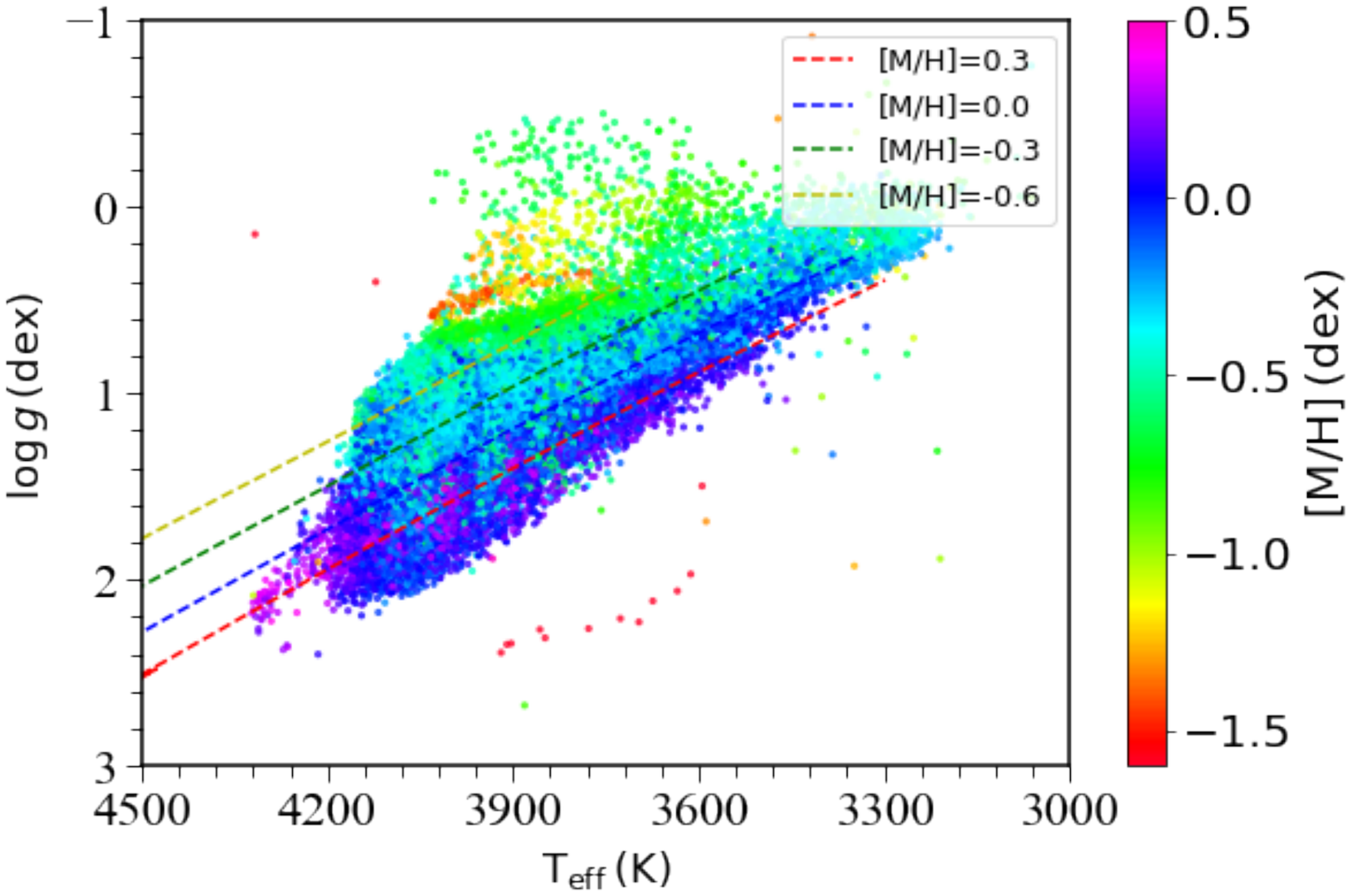}
\caption{The distributions in the HR diagram of the training set and the prediction M giants with SNR $>$ 50
are showed color-coded by the metallicity [M/H] 
in the left and right panels, respectively. In the
left panel, the stellar parameters of 3670 M giant stars are from APOGEE.
In the right panel, the stellar parameters of the M giant sample with SNR $>$ 50 are predicted by
SLAM. The dashed lines in both panels represent the isochrones 
from PARSEC model with the same age 3 Gyr but different metallicities, 
i.e.  -0.6, -0.3, 0 and 0.3 dex for the yellow, green, blue and red lines, respectively.
}\label{fig:Train_label}
\end{figure}

The excluded 803 stars will be used 
in the subsection \ref{sect:self} to verify the self consistency 
of the trained model in deriving the stellar parameters.
Figure \ref{fig:Train_label} shows 
the Hertzsprung Russell diagram (HRD) of the 3670 common M giants stars,
which are color-coded by the metallicity [M/H]. 
We find that all the training samples are located in the ranges of
-1.5 $<$ [M/H] $<$ 0.5 dex, 3200 $<$ ${T_{\mathrm{eff}}}$ $<$ 4300 k, 
-0.4 $<$ log$\,g$ $<$ 2.5 dex. For comparison, we also represent 
the isochrones from the PAdova and TRieste Stellar Evolution Code
\citep[PARSEC;][]{Bressan-2012} with the dashed lines of the same age of 3 Gyr and
different metallicities of 0.3, 0, -0.3 and -0.6 dex. As showed in the later results, 
this is reasonable for our sample, majority of which are the thin disk members.
Statistically speaking, the distribution of the training stellar 
labels are consistent with the stellar evolution model and the M giant stars are mainly the metal rich stars.

\section{Method and Results}\label{sect:Method}

\subsection{Radial Velocity}\label{sect:D_radial_velocity}

The radial velocity is derived by using cross correlation based method \emph{laspec} \citep{Zhang-2021}. It is applied on the spectra of M giant stars with wavelength between 8000 to 8950 \AA \, from LAMOST DR9, where Ca II triplet is included in this band.
The results can be evaluated by the following equation,
\begin{equation}\label{equ:CCF}
{\rm CCF}(v|F,G) = \frac{{\rm Cov}(F, G(v))}{\sqrt{{\rm Var}(F){\rm Var}(G(v))}}
\end{equation}
where $F$ is the normalized observed spectrum of the given M giant star, while $G$
is the synthetic spectra from ATLAS9 \citep{ATLAS92018A&A...618A..25A}, also with 
wavelength between 8000 to 8950 \AA  \, with a shift caused by the radial velocity. It is noteworthy that the spectra used in subsection \ref{sect:Para} have been shifted back to the rest frame by the radial velocity.

\begin{figure*}
\centering
\includegraphics[width=0.38\textwidth, trim=0.cm 0.0cm 2.5cm 1.cm, clip]{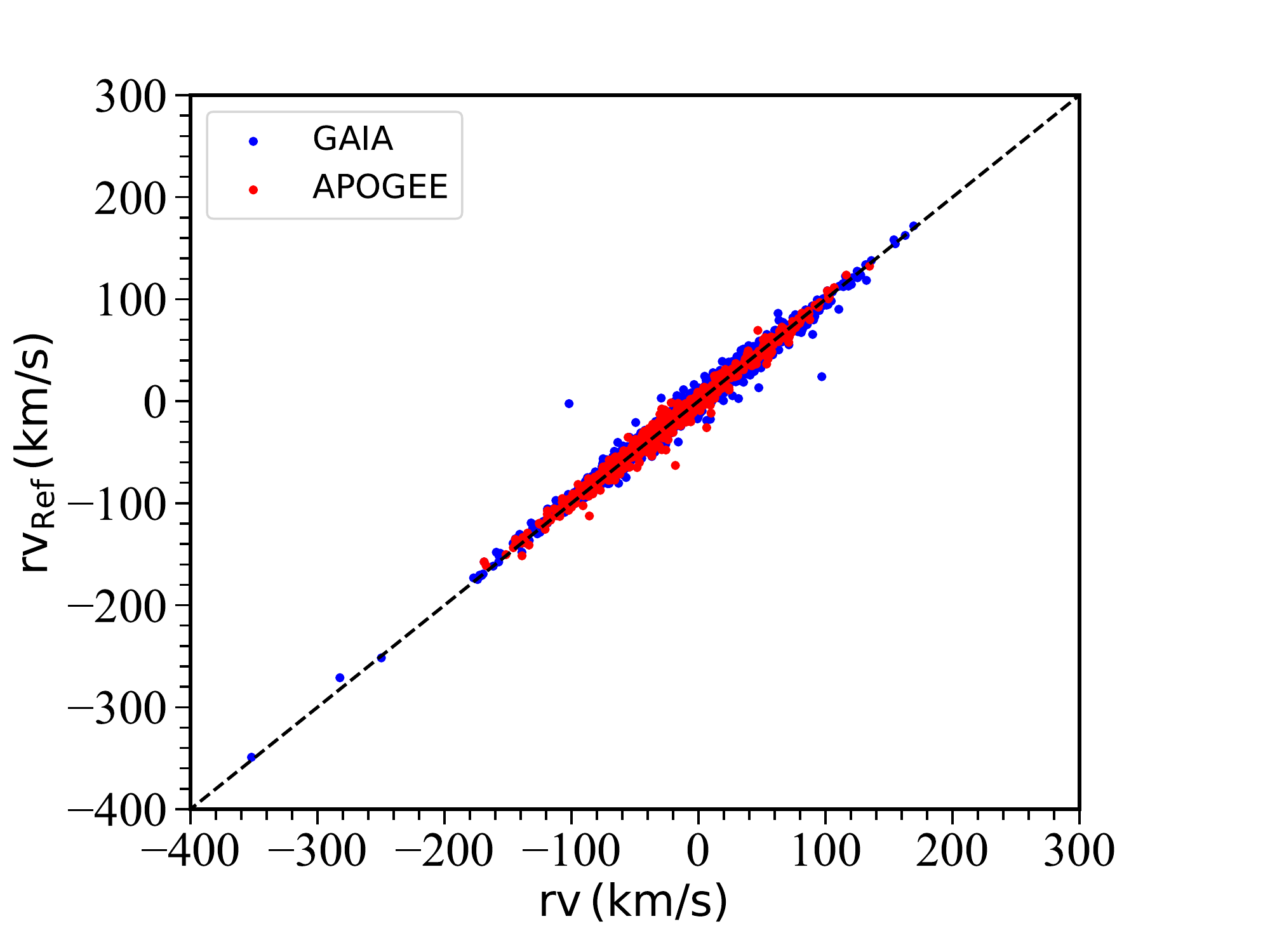}
\includegraphics[width=0.38\textwidth, trim=0.cm 0.0cm 2.8cm 1.cm,clip]{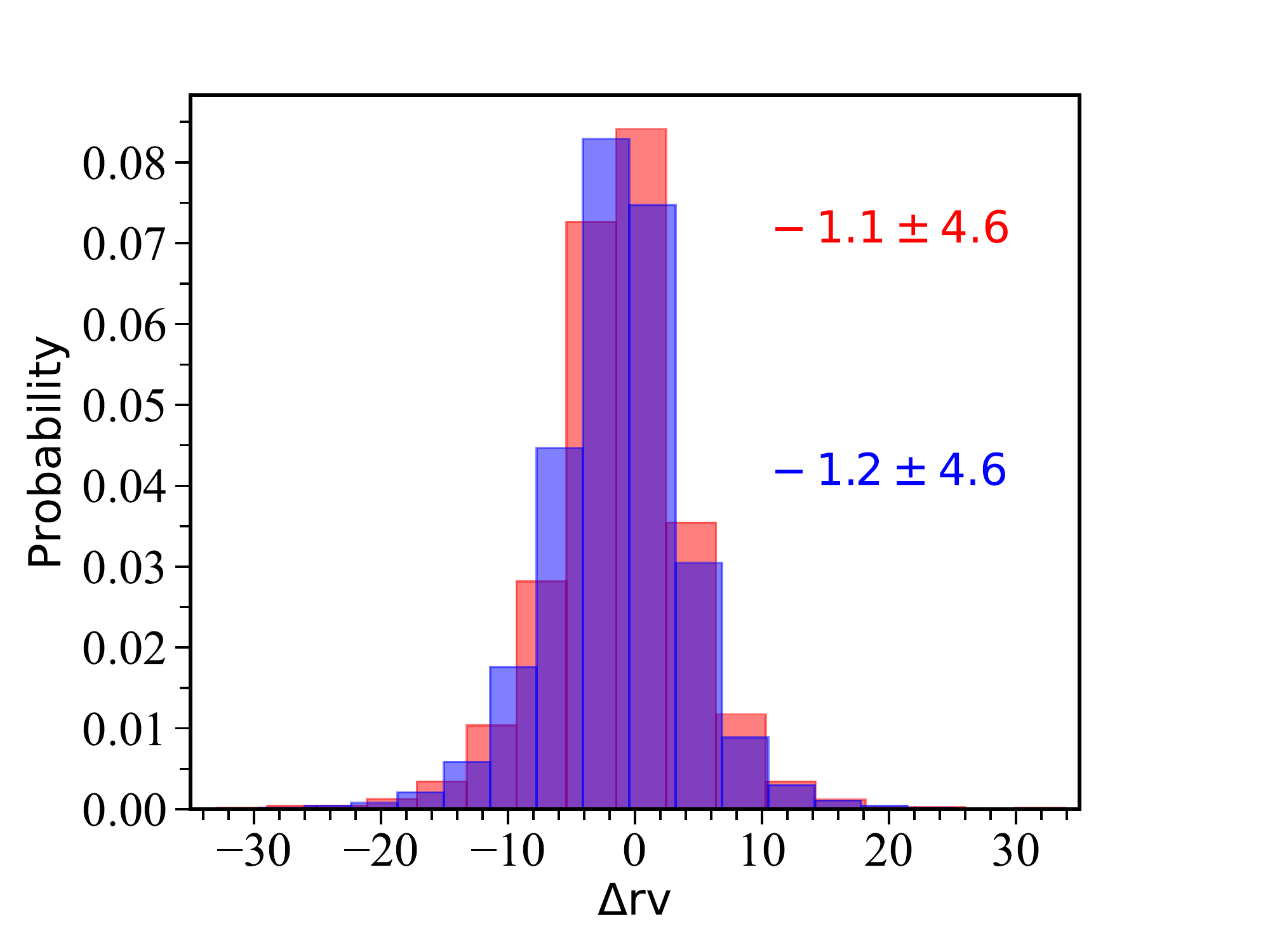}
\caption{Compared the radial velocity from this work with that 
in Gaia DR3 (32787 stars) and APOGEE DR17 (4888 stars), as displayed 
by the red and blue colors in the left panel, respectively. The right
panel displays the corresponding histograms of radial velocity difference
$\rm \Delta rv$, where rv is the radial velocity derived 
in this work, and rv$\rm_{Ref}$ is the radial velocity from Gaia DR3 (blue)
and APOGEE DR17 (red).}\label{fig:rv}
\end{figure*}

To evaluate the results, we cross match our sample with Gaia DR3
\citep[Gaia DR3;][]{Katz-2022,Gaia-2022}
and APOGEE DR17 \citep{Abdurro-2022} to 
obtain two datasets of common stars. There are 32787  and 4888 common stars 
with Gaia DR3 and APOGEE DR17, respectively. Then
the comparison of our results with Gaia DR3 and APOGEE DR17 
are represented by the blue and red symbols in the Figure \ref{fig:rv}, respectively. 
It shows a good agreement between our results with that from  Gaia DR3 and 
APOGEE DR17. The histogram of the 
radial velocity difference $\rm \Delta rv=rv-rv\rm_{Ref}$ are showed in the right panel.
The system offset and scatter values of $\rm \Delta rv$ are around $\sim$ 1 km s$^{-1}$
and 4.6 km s$^{-1}$, respectively, for both comparison results. It indicates 
that the radial velocity of M giants derived form Ca II Triplet lines is 
consistent with that of Gaia DR3 and APOGEE DR17. 

\subsection{Stellar Parameter} \label{sect:Para}
The Stellar LAbel Machine (SLAM) is a data-driven method based
on support vector regression (SVR) \citep{Zhang-2020}. 
SVR is a robust nonlinear regression which has been applied in many
fields of astronomy \citep{Liu-2012,Liu-2015}. 
SVR has been particularly widely 
used in spectral data analysis \citep{Li-2014, Liu-2014,Bu-2015}. 
This method has been proved to have good performance in determining 
stellar atmospheric parameters from spectra \citep{Zhang-2020,Li-2021,Guo-2021}. 
In this work, we also adopt SLAM to derive the atmospheric parameters 
of the M giants.

\subsubsection{Stellar label model training}\label{sect:Model_tra}
In SLAM, the radial basis function (RBF) is adopted as the kernel of SVR. 
The hyperparameters $C$, $\varepsilon$ and $\gamma$ of 
SLAM represent the penalty level, tube radius 
and the width of the RBF kernel, respectively. These three hyperparameters
can be automatically determined for each pixel through the training set. 

$\boldsymbol{\vec{\theta}}_i$ is denoted as the stellar label vector of the $i$th star in the training set. \emph {f$_j$}($\boldsymbol{\vec{\theta}}_i$) 
and $f_{i,j}$ are defined as the $j$th pixel of the training spectrum and model output spectrum corresponding to the the stars with stellar label 
vector $\boldsymbol{\vec{\theta}}_i$. The mean squared error (MSE) and 
median deviation (MD) of $j$th pixel can be evaluated with a specific 
set of hyperparameters, described by Equations (\ref{eq:MSE}) and (\ref{eq:MD})

\begin{equation}\label{eq:MSE}
MSE_\emph{j}=\frac{1}{m}\sum_{i=1}^m\big[\emph{f$_j$}\big(\boldsymbol{\vec{\theta}}_i\big)-\emph{f$_ {i,j}$}\big]^2
\end{equation}
\begin{equation}\label{eq:MD}
MD_\emph{j}=\frac{1}{m}\sum_{i=1}^m\big[\emph{f$_j$}\big(\boldsymbol{\vec{\theta}}_i\big)-\emph{f$_{i,j}$}\big]
\end{equation}
\\
Theoretically, the smaller MSE and MD are, the better fitting is. However, 
we probably get an overfitted model if we train the SLAM model by whole 
training set, i.e., the $\rm MSE_\emph{j}$ and $\rm MD_\emph{j}$ are all 
equal to 0. \citet{Zhang-2020} used the k-fold cross-validated MSE (CV MSE) 
and $k$-flod cross-validated MD (CV MD) to measure MSE$_{j}$ and MD$\rm_{j}$ to
avoid getting an overfitted trained model. Namely, the training set is randomly 
divided into $k$ subsets, where $k$ is set to be 10 in this work. 
The \emph {f$_j$}($\boldsymbol{\vec{\theta}}_i$) is predicted by
the model which trained
by the other $k-1$ subsets of the training set. After looping through
all the hyperparameters sets specified, the best set of hyperparameters 
can be determined for $j$th pixel by searching for the lowest 
CV MSE$_{j}$. The best model can be obtained for each pixel
by doing pixel-to-pixel. In this work, we train SLAM model with 3670 low-resolution spectra from LAMOST 
of the training sample and their corresponding 
stellar labels from APOGEE.

\subsubsection{Prediction of stellar labels} \label{sect:Pre_data} 
Using the Bayesian formula, the posterior probability density function of 
stellar label vector for a given observed spectrum is displayed in Equation  (\ref{eq:PDF}).

\begin{equation}\label{eq:PDF}
p\big(\boldsymbol{\vec{\theta}}\mid\boldsymbol{\vec{f}}_{obs}\big)\propto p\big(\boldsymbol{\vec{\theta}}\big)\prod_{j=1}^n p\big(f_{j,obs}\mid\boldsymbol{\vec{\theta}}\big)
\end{equation}
where $\boldsymbol{\vec{\theta}}$ is the stellar label vector,
$\boldsymbol{\vec{f}}_{obs}$ and \emph{f$_{j,obs}$} represent the
normalized observed spectrum vector and the normalized flux of \emph{j}th 
pixel of the observed spectrum, respectively. $p(\boldsymbol{\vec{\theta}})$ 
is the prior of stellar label vector $\boldsymbol{\vec{\theta}}$,
$p(f_{j,obs}\mid\boldsymbol{\vec{\theta}})$ is the likelihood of observed spectrum flux of
\emph{j}th pixel with a given
stellar label vetor $\boldsymbol{\vec{\theta}}$. By maximizing the 
posterior probability $ p(\boldsymbol{\vec{\theta}}\mid\boldsymbol{\vec {f}}_{obs})$, 
the stellar labels can be easily measured with a 
Gaussian likelihood adopted. Then the logarithmic form of likelihood is 
described by the following equation,

\begin{equation}\label{eq:likeli}
\begin{split}
\ln p\big(\boldsymbol{\vec{\theta}}\mid\boldsymbol{\vec{f}}_{obs}\big)=&-\frac{1}{2}\sum_{j=1}^n\times\frac{\big[f_{j,obs}-f_j\big(\boldsymbol{\vec{\theta}}\big)\big]^2}{\sigma_{j,obs}^2+\sigma_j\big(\boldsymbol{\vec{\theta}}\big)^2}\\
&-\frac{1}{2}\sum_{j=1}^n\times\ln\big[2\pi\big(\sigma_{j,obs}^2+\sigma_j\big(\boldsymbol{\vec{\theta}}\big)^2 \big)\big]
\end{split}
\end{equation}
\\
where $f_{j}\big(\boldsymbol{\vec{\theta}}\big)$ and $\sigma_j(\boldsymbol{\vec{\theta}})$ 
are the model output spectrum and the uncertainty of \emph{j}th pixel 
corresponding to stellar label $\boldsymbol{\vec{\theta}}$, 
respectively. $f_{j,obs}$ is the $j$th pixel of the normalized observed 
spectrum, and $\sigma_{j,obs}$ is the uncertainty of \emph{j}th pixel of normalized observed spectrum.

The trained model as described in subsection \ref{sect:Model_tra} is applied on the spectra
of all the M giant stars to obtain their atmospheric parameters.
The right panel of Figure \ref{fig:Train_label} 
illustrates the distribution of the effective temperature $T_{\mathrm{eff}}$ versus 
surface gravity $\mathrm{log}\,g$ of the prediction M giant stars with SNR $>$ 50, accounting for 80\% of the whole M giant stars. The color is coded by the metallicity [M/H]. 
It exhibits a similar pattern with that of the training sample as 
showed in the left panel of Figure \ref{fig:Train_label}. The same isochrones are also represented by the dashed lines with those
in the first panel of Figure \ref{fig:Train_label}.

\subsubsection{Stellar label uncertainty}\label{sect:uncer}
Similar to the CV MSE and CV MD of spectrum as described in \ref{sect:Model_tra}, 
Equations (\ref{eq:bias}) and (\ref{eq:scatter}) describe the cross-validate scatter
(CV\_scatter) and cross-validate bias (CV\_bias) of stellar labels, respectively.
Obviously, if the predicted stars have the known stellar label, 
the CV\_scatter and CV\_bias can be measured. They can be regarded as the
standard deviation and average deviation of the stellar labels, respectively,
to describe the precision of the stellar parameters determined from SLAM
model. In principle, the smaller CV\_bias and CV\_scatter indicate a 
better trained model.

\begin{equation}\label{eq:bias}
CV\_bias=\frac{1}{m}\sum_{i=1}^m\big(\boldsymbol{\vec{\theta}}_{i,SLAM}-\boldsymbol{\vec{\theta}}_{i}\big)
\end{equation}

\begin{equation}\label{eq:scatter}
CV\_scatter=\frac{1}{m}\sqrt{\sum_{i=1}^m\big(\boldsymbol{\vec{\theta}}_{i,SLAM}-\boldsymbol{\vec{\theta}}_{i}\big)^2}
\end{equation}
where $\boldsymbol{\vec{\theta}}_{i,SLAM}$ is the stellar label vector
from the model prediction, and $\boldsymbol{\vec{\theta}_{i}}$ is the 
corresponding true stellar label vector. 

\begin{figure*}
\centering
\includegraphics[width=0.4\textwidth]{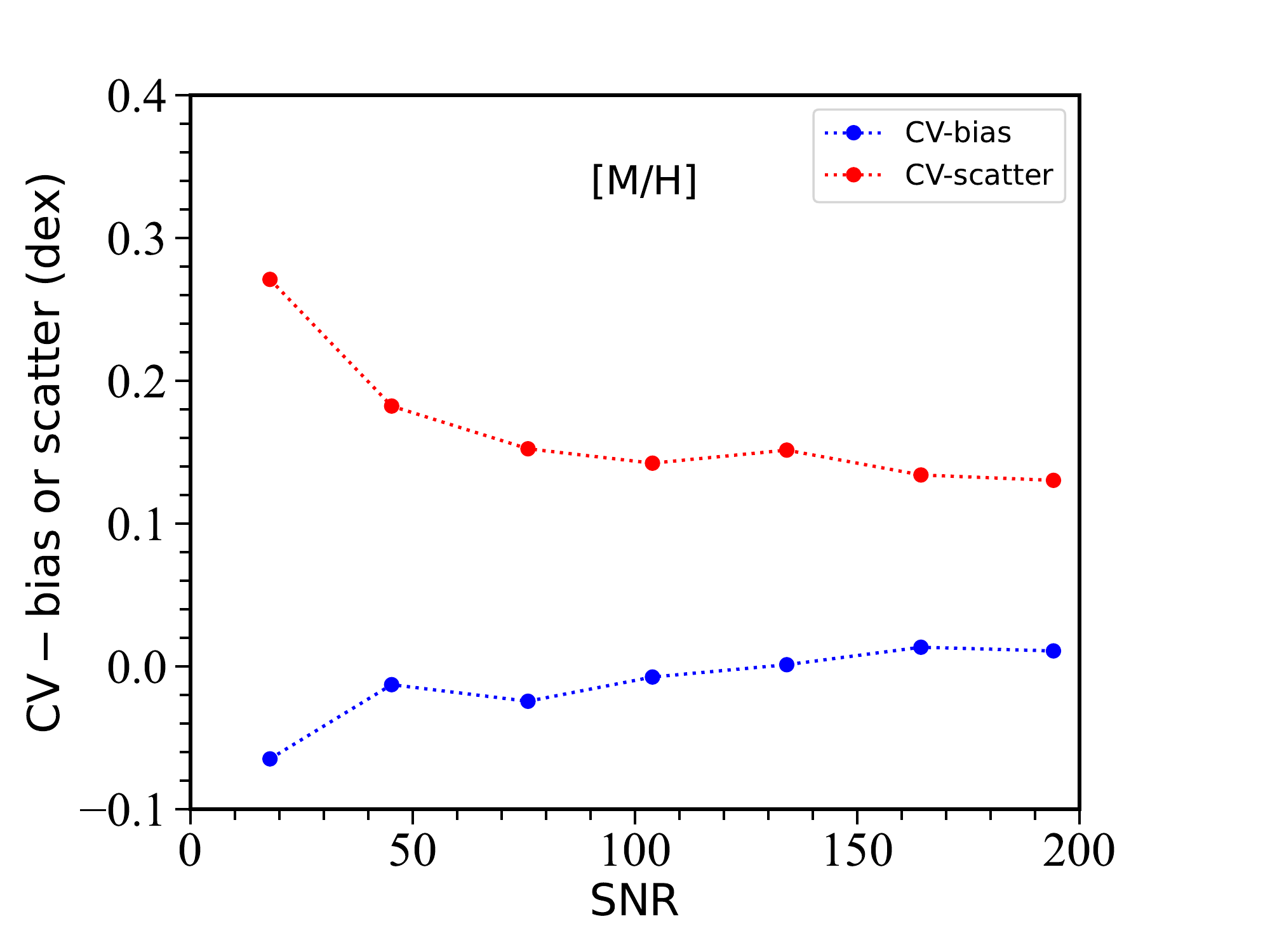}
\includegraphics[width=0.4\textwidth]{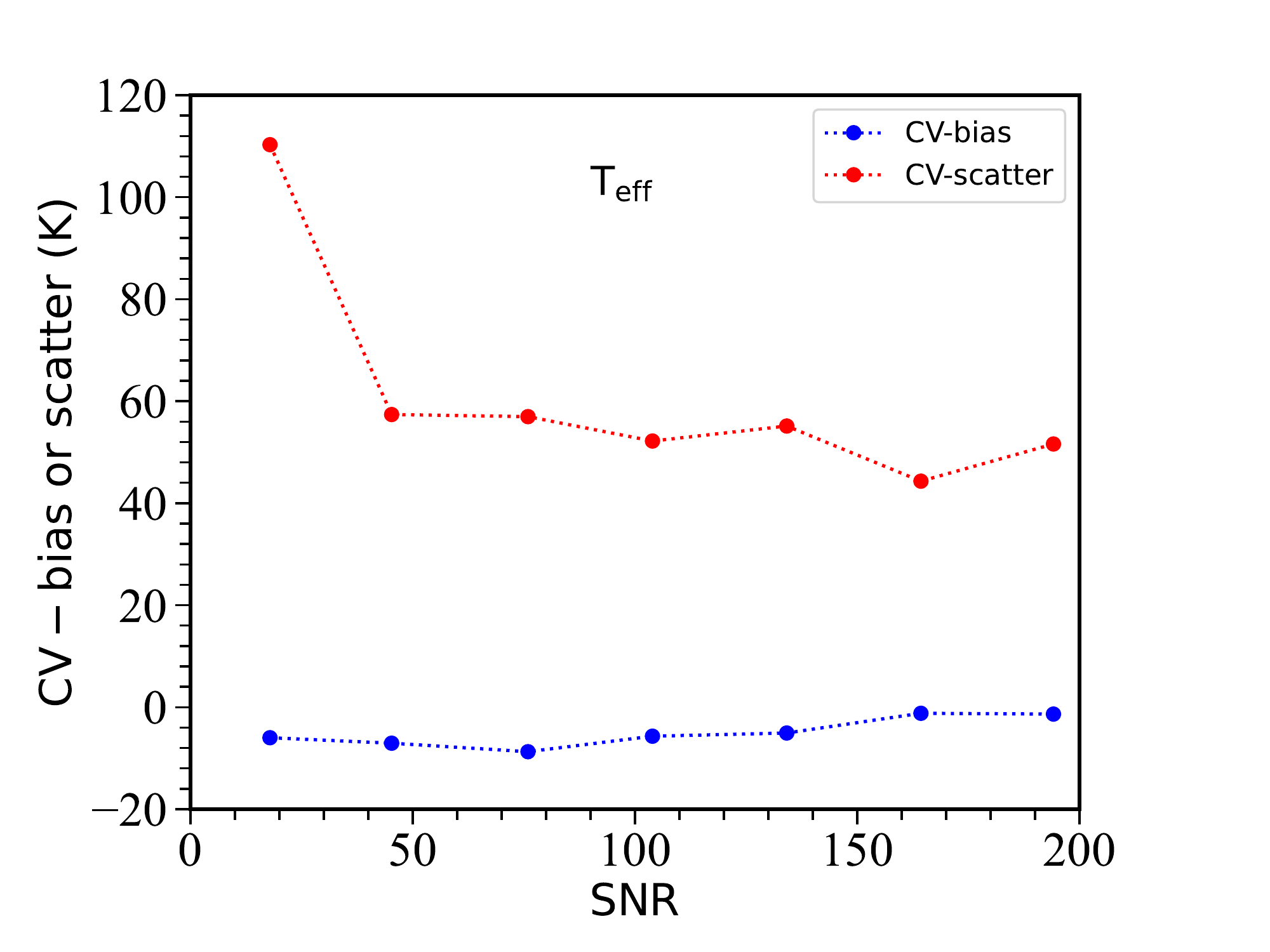}
\includegraphics[width=0.4\textwidth]{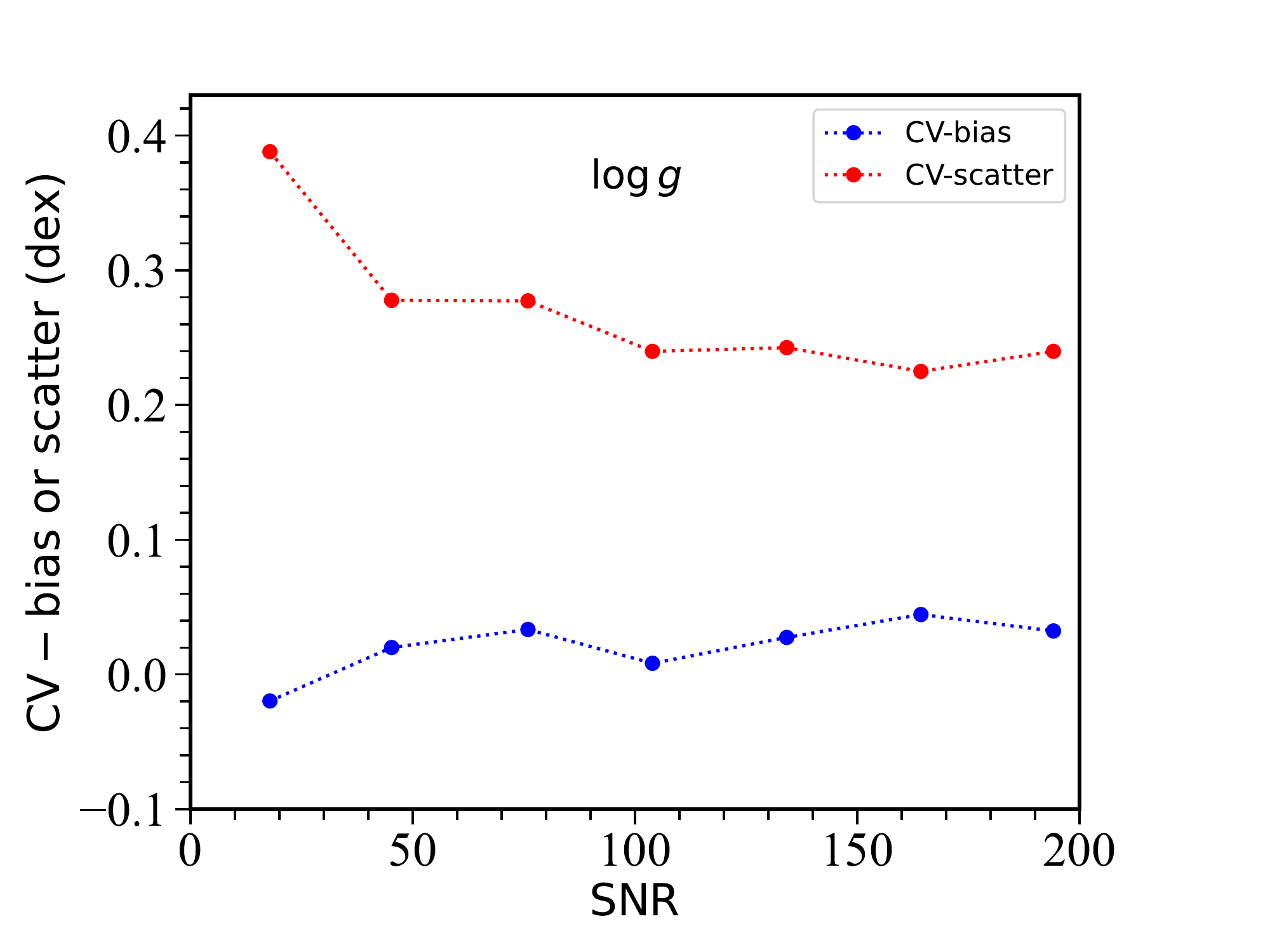}
\includegraphics[width=0.4\textwidth]{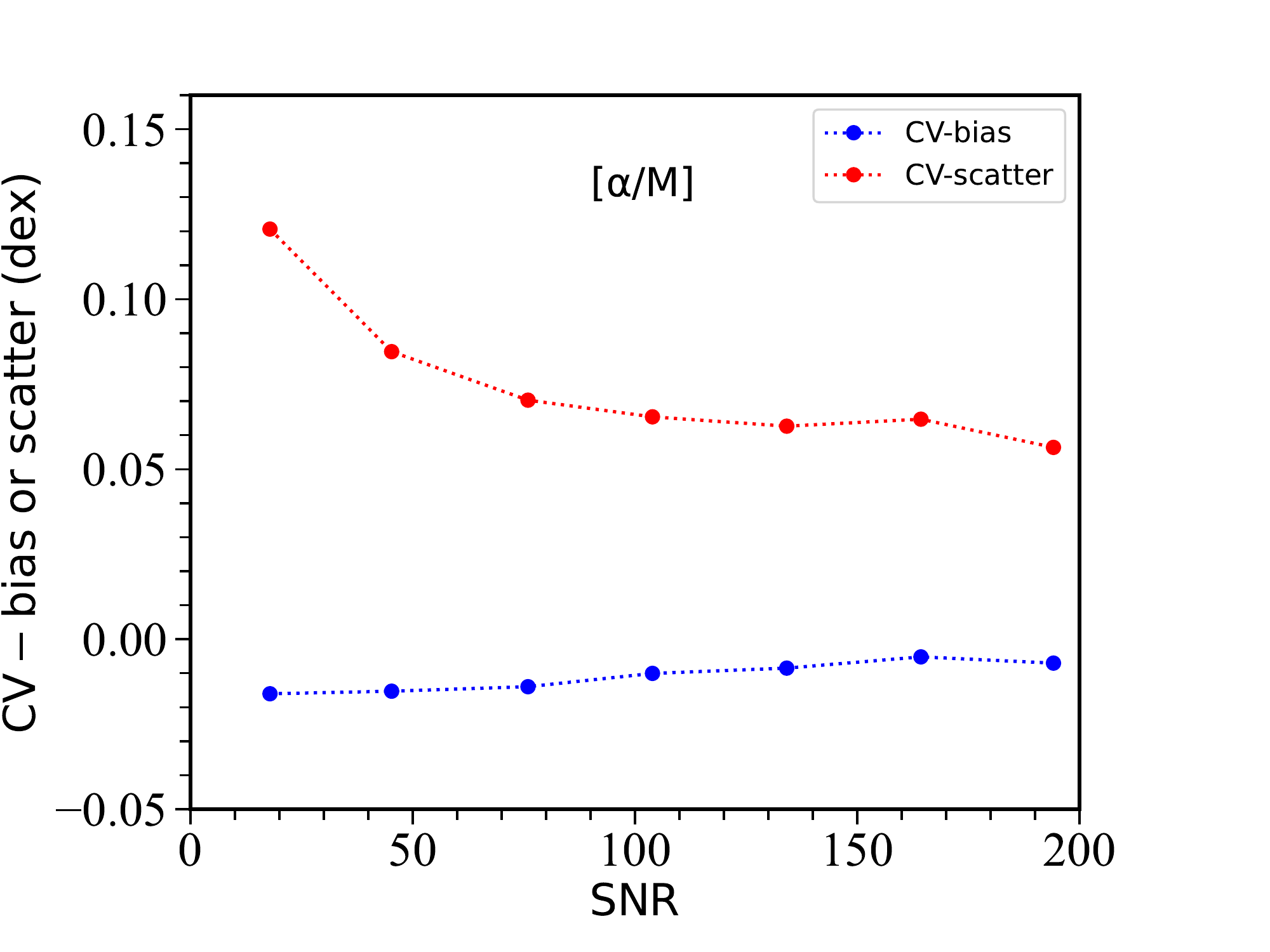}

\caption{Four panels exhibit the CV scatter and bias as a function of SNR for 
[M/H], $T_{\mathrm{eff}}$, log$\,g$ and [$\alpha$/M], respectively. 
The red and blue dotted lines indicate the CV scatter and CV bias
in different SNR bins, respectively. }\label{fig:cv}
\end{figure*}

Figure \ref{fig:cv} shows the distributions of CV bias (blue) and scatter (red) of the four parameters
versus the SNR of spectra, i.e. the metallicity [M/H], the effective temperature $T_{\mathrm{eff}}$,
the surface gravity  $\mathrm{log}\,g$ and  the alpha abundance [$\alpha$/M]. 
It is obvious that the CV scatters of these four parameters decrease with increasing SNR,
e.g. from 0.27 dex, 110 K, 0.39 dex, 0.12 dex at SNR=17 to 0.16 dex, 
57 K, 0.25 dex and 0.06 dex at SNR=100 for 
 [M/H], $T_{\mathrm{eff}}$, log$\,g$ and [$\alpha$/M], respectively.
 In other words, the precision of these 
four parameters determined form SLAM model can reach to 0.16 dex, 57 K, 0.25 dex and 0.06 dex
at SNR $>$ 100, respectively. The mean values of CV bias are -0.01 dex, -5 K, 0.02 dex and 
-0.01 dex for these four parameters, respectively, which means that the predicted parameters by 
SLAM are in good agreement with the true stellar labels given by APOGEE. 

\subsubsection{Stellar labels self-consistent}\label{sect:self}
\begin{figure*}
\centering
\includegraphics[width=0.38\textwidth]{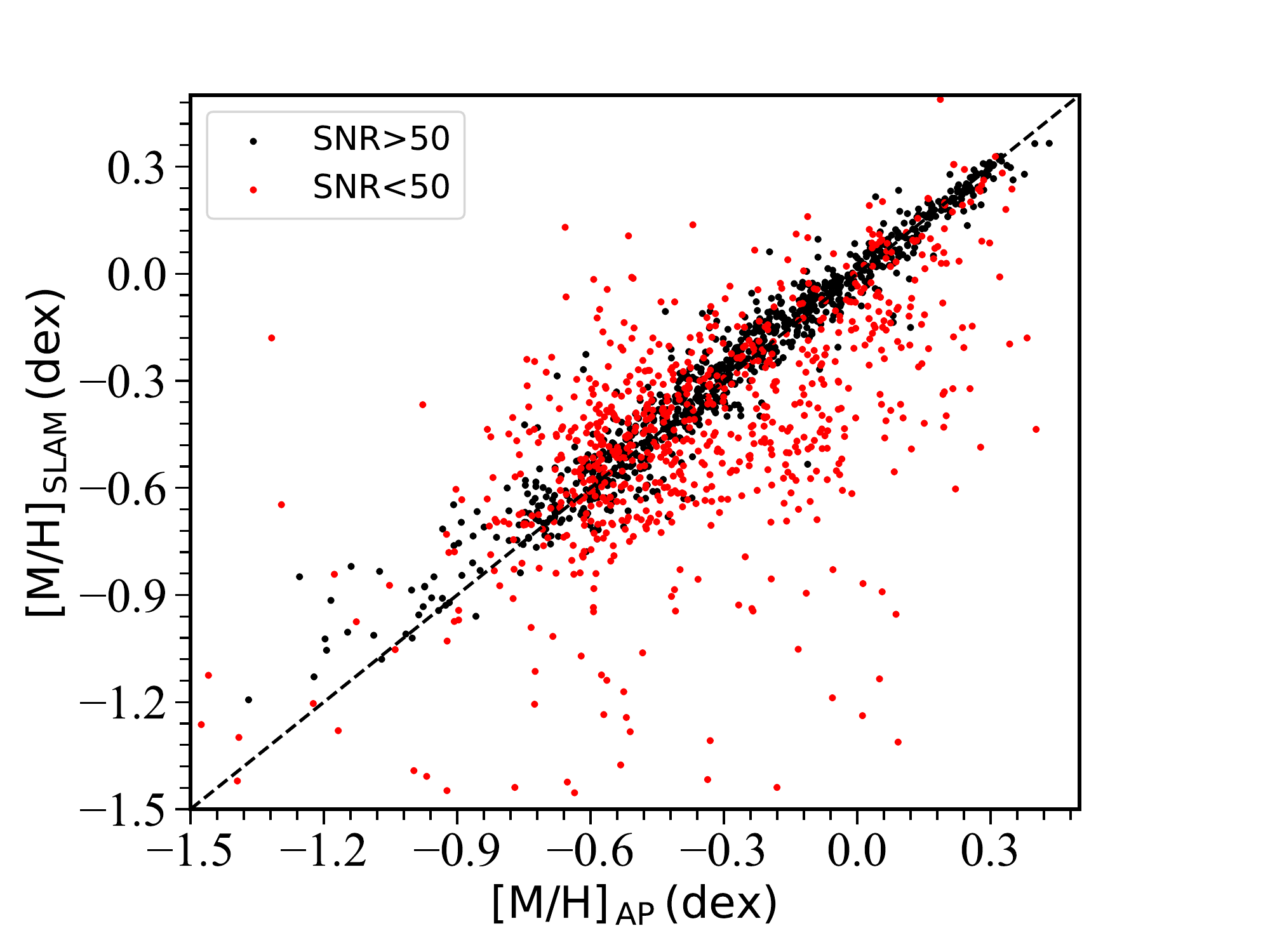}
\includegraphics[width=0.38\textwidth]{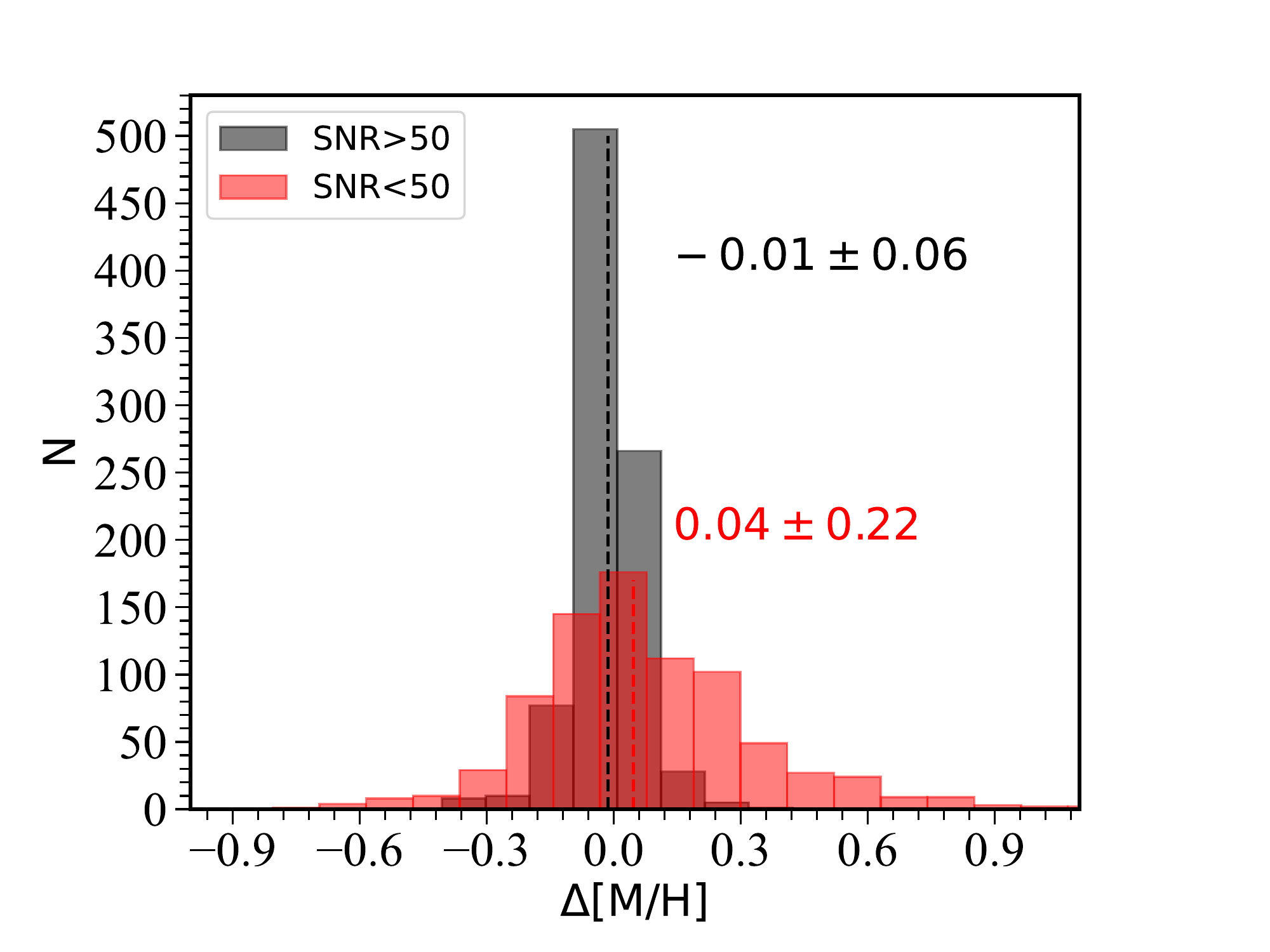}
\includegraphics[width=0.38\textwidth]{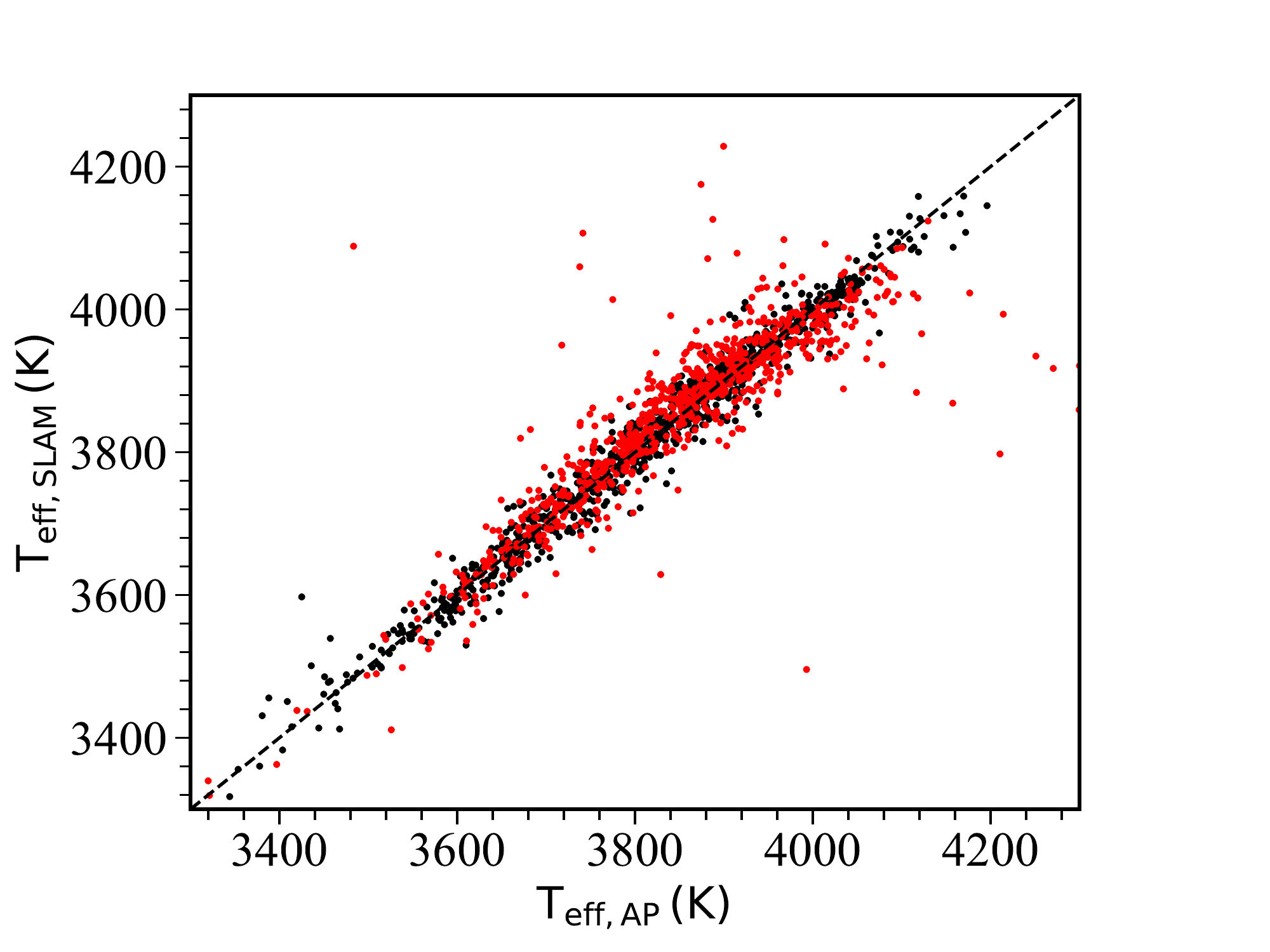}
\includegraphics[width=0.38\textwidth]{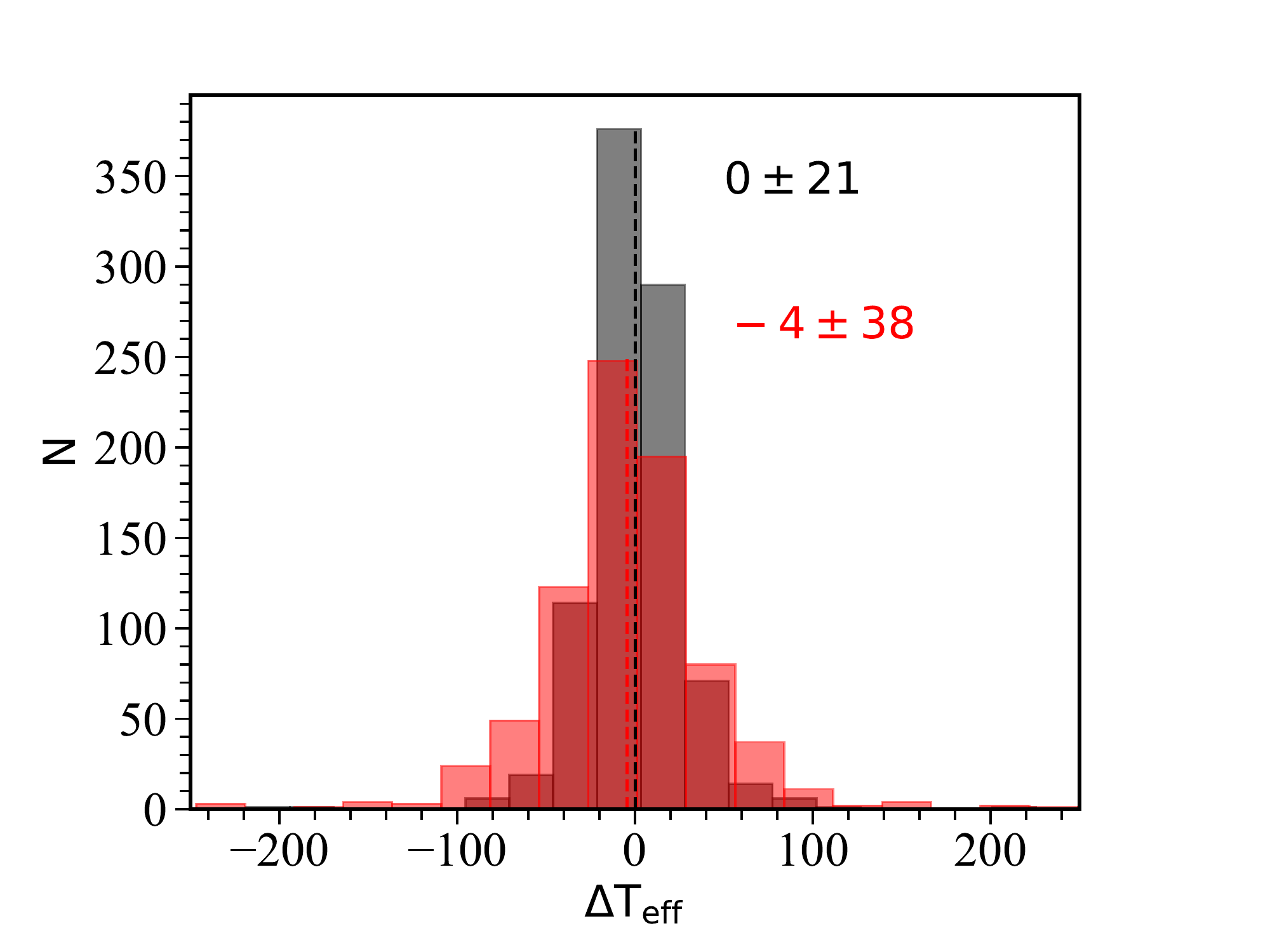}
\includegraphics[width=0.38\textwidth]{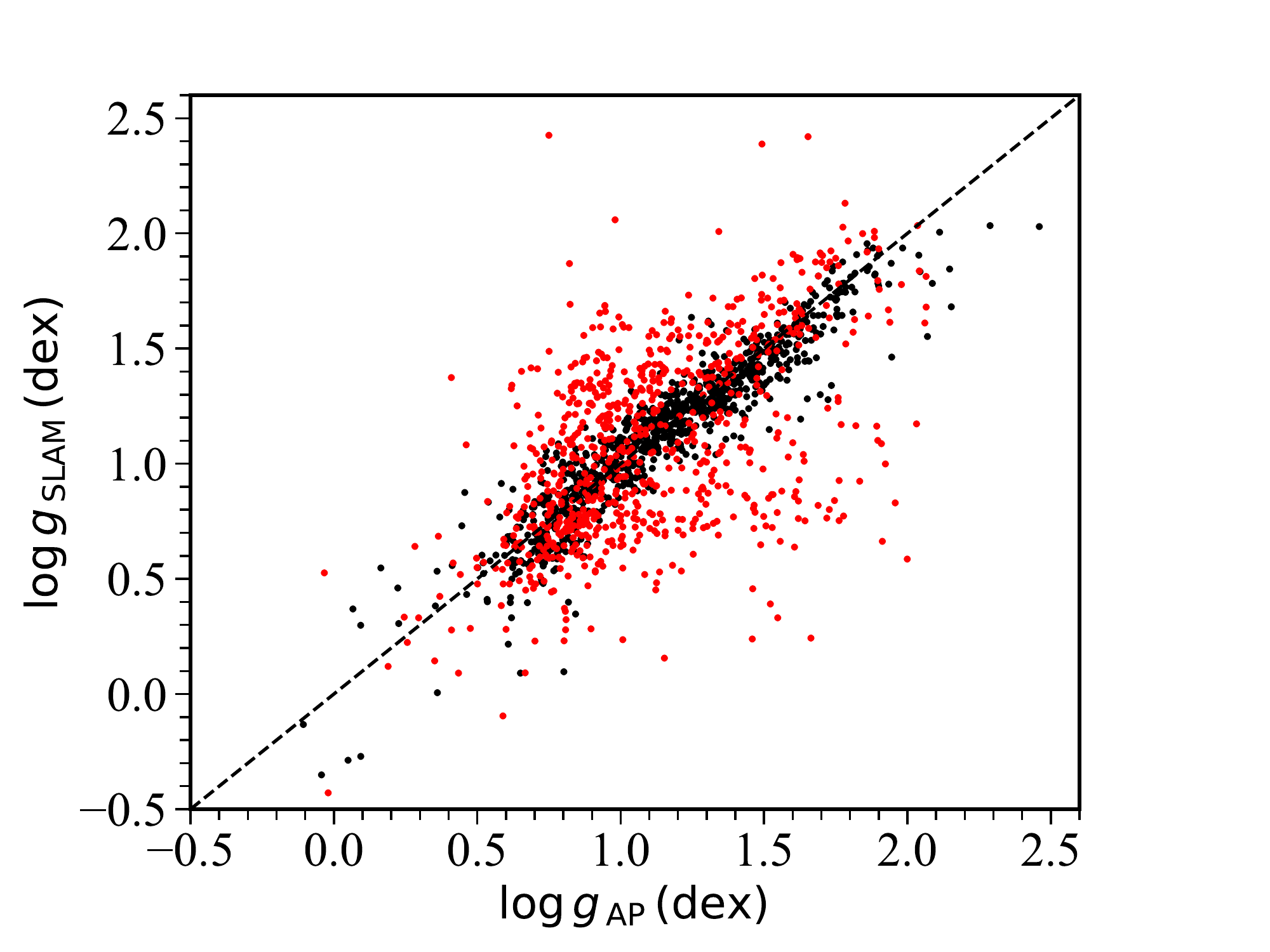}
\includegraphics[width=0.385\textwidth]{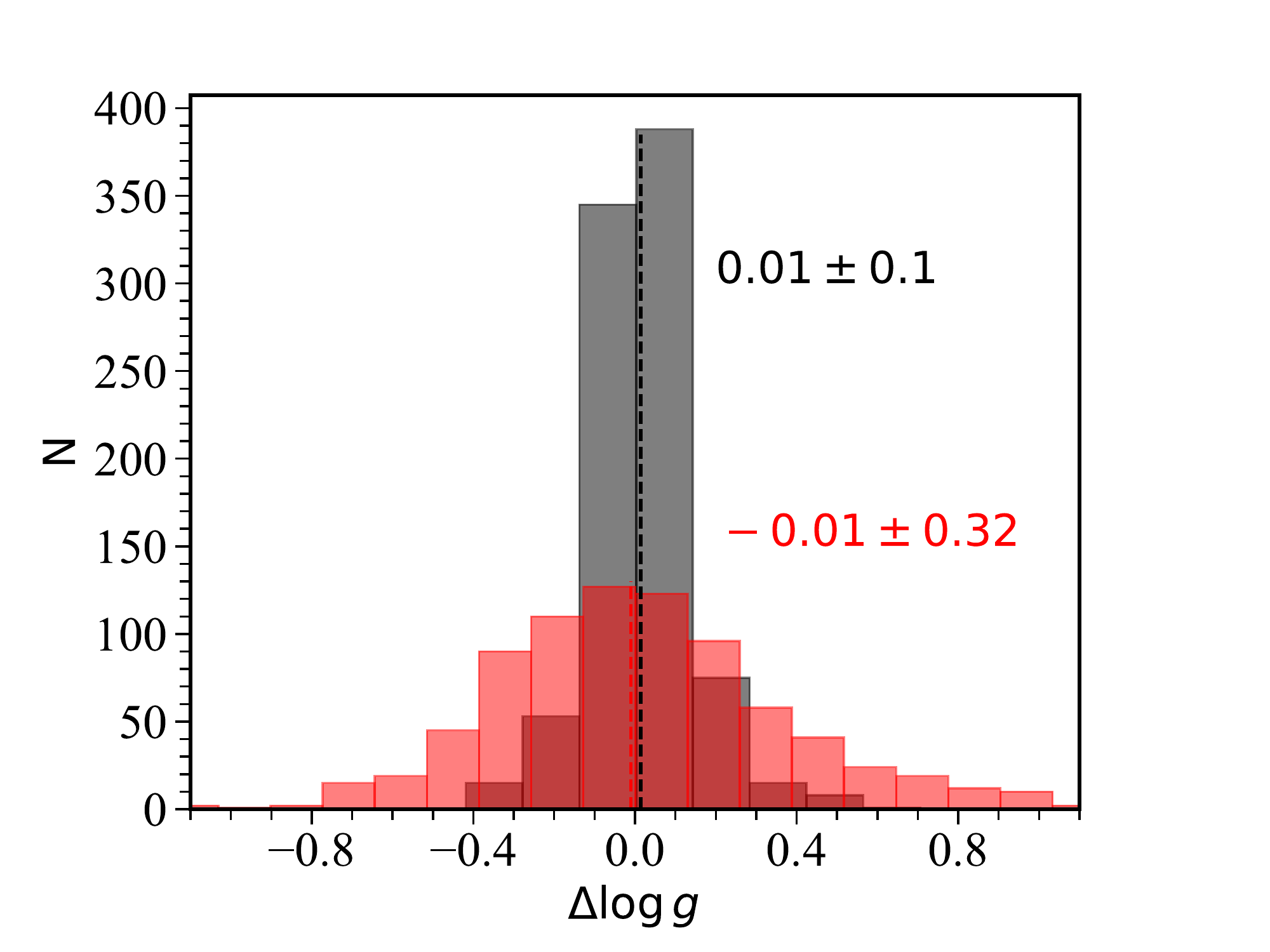}
\includegraphics[width=0.38\textwidth]{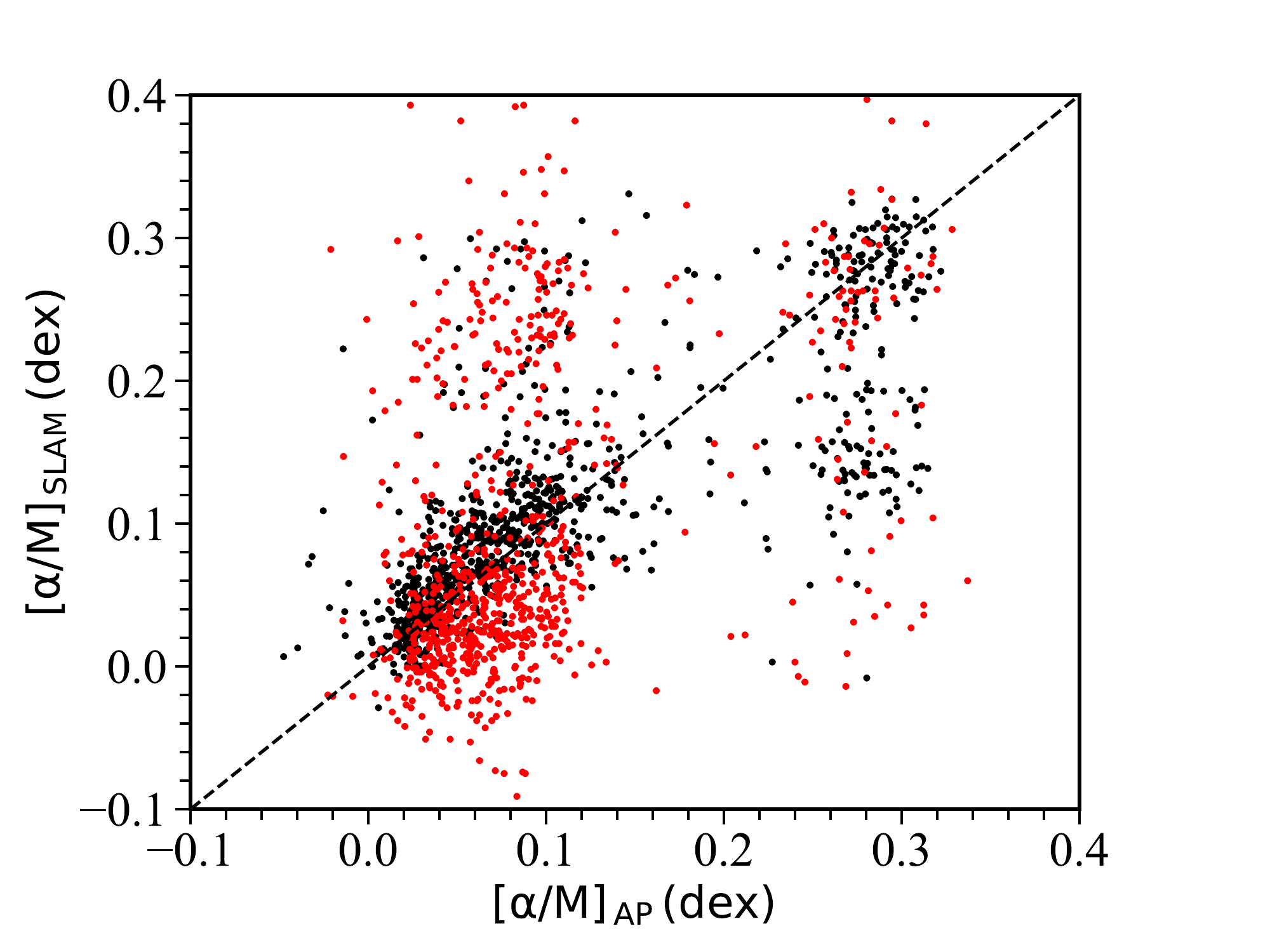}
\includegraphics[width=0.38\textwidth]{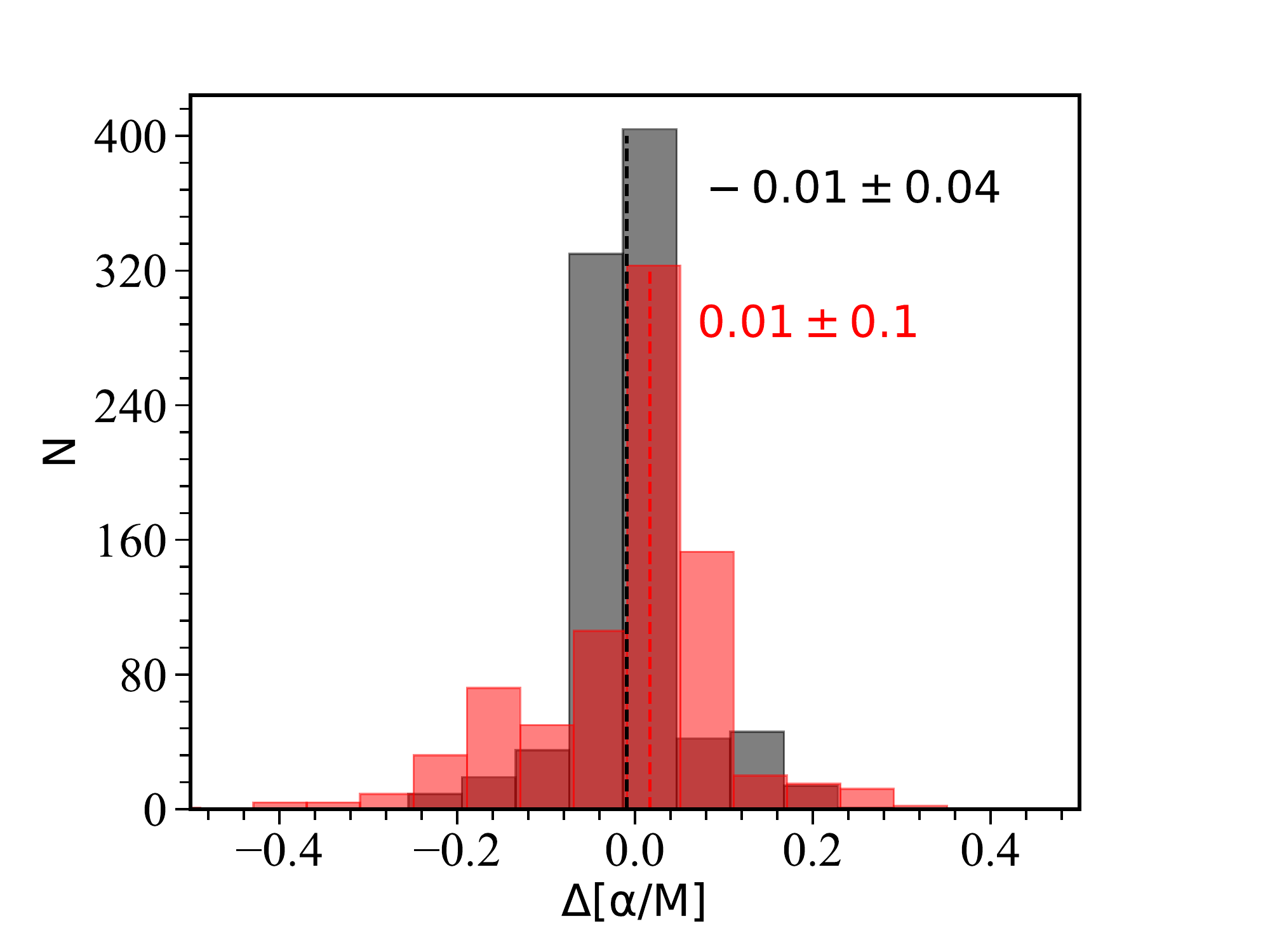}

\caption{Four panels on the left display the comparison of stellar 
labels between the true values $\rm X_{AP}$ and model prediction values $\rm X_{SLAM}$, 
where X can be [M/H], $T_{\mathrm{eff}}$, log$\,g$ and [$\alpha$/M]. 
The black and red dots
represent the stars with SNR$>50$ and SNR$<50$, respectively. 
The black dashed line is the
one to one line in these four panels. The histograms
of $\rm \Delta X$ (= $\rm X_{AP}$-$\rm X_{SLAM}$) as
exhibited in the right four panels. The black and 
red histograms show the $\rm \Delta X$ of
stars with SNR $>$ 50 and stars with SNR $<$ 50, respectively. The mean values of 
$\rm \Delta X$ of two subsamples as marked by the black and 
red dashed lines.}\label{fig:Train}
\end{figure*}

In order to verify the self consistency of stellar labels determined from SLAM model. 
The training set is randomly split into two subsets, a training set consists of 2770 M giants, 
the remaining 900 M giants are used as the test set. Besides, the 803 M giant 
stars mentioned in subsection \ref{sect:Train_set} are also used as the test
set. Figure \ref{fig:Train} displayed the comparison of four stellar labels between
the true values and the model prediction values. In the left four panels, the black 
and red dots display the $\rm X_{AP}$ vs. $\rm X_{SLAM}$ of 900 stars with SNR$>50$ 
and 803 stars with SNR$<50$ in the test set, respectively, where X can be [M/H], 
$T_{\mathrm{eff}}$, log$\,g$ and [$\alpha$/M], respectively. It is obviously that 
the model prediction stellar labels are agreement with the true stellar labels. 
However, the distribution between $\rm X_{AP}$ and $\rm X_{SLAM}$ of stars with 
low SNR (red dots) is more dispersed than that of stars with high SNR (black dots). 
As shown in the four right panels, the red and black histograms exhibit the distribution 
of $\rm \Delta X$ = $\rm X_{AP}$-$\rm X_{SLAM}$ of stars with SNR $>$ 50 and stars with 
SNR $<$ 50. The mean values of $\rm \Delta X$ of two subsamples are close to 0, but the
scatter values of stars with low SNR is larger than that of stars with higher SNR. Because
the spectra with low signal-to-noise
ratio are more difficult to obtain high-precision stellar parameters.

\subsubsection{Stellar labels validation}\label{sect:validation}
\begin{figure*}
\centering
\includegraphics[width=0.38\textwidth, trim=0.cm 0.0cm 3.0cm 1.cm, clip]{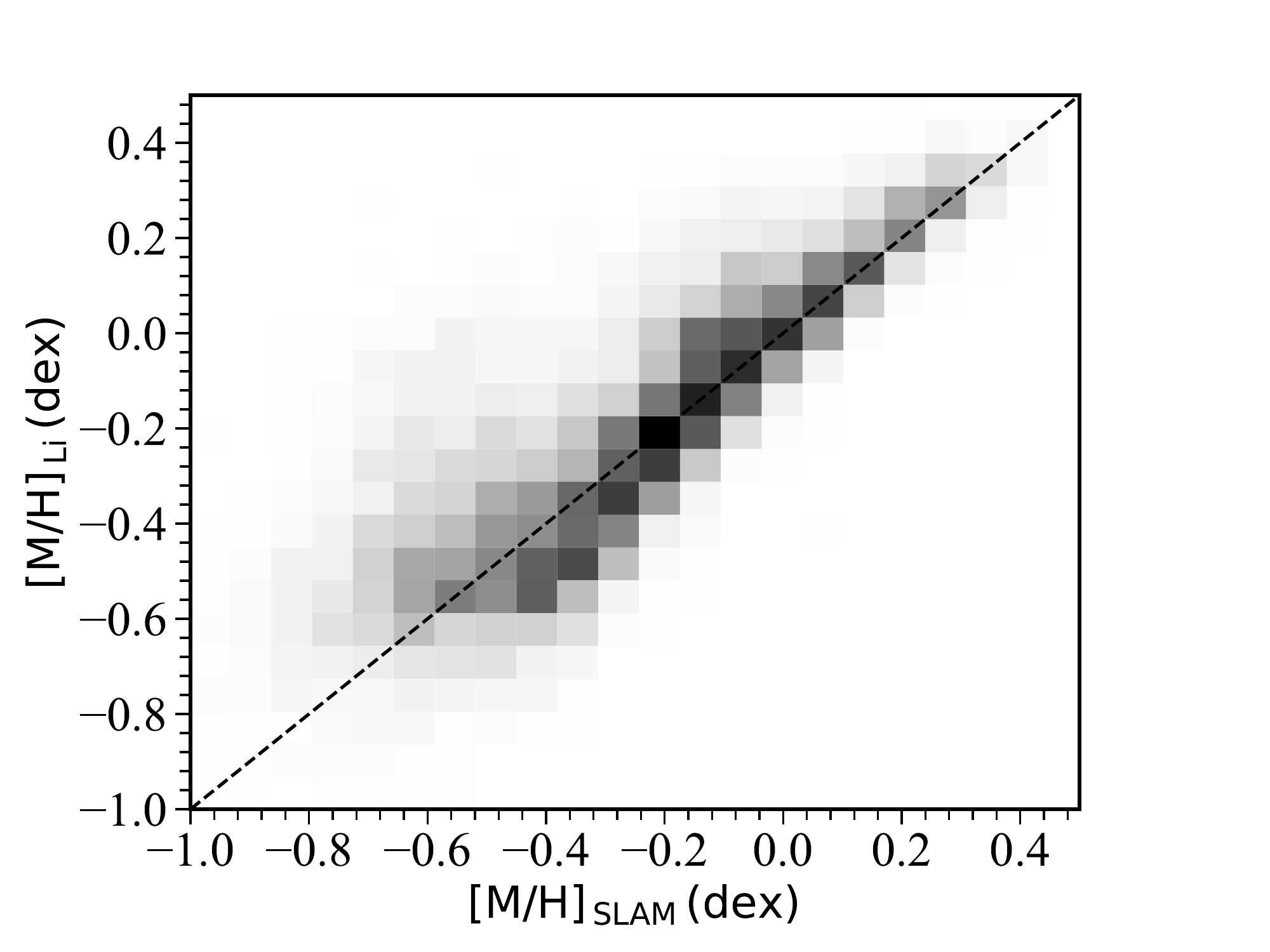}
\includegraphics[width=0.38\textwidth, trim=0.cm 0.0cm 3.0cm 1.cm,clip]{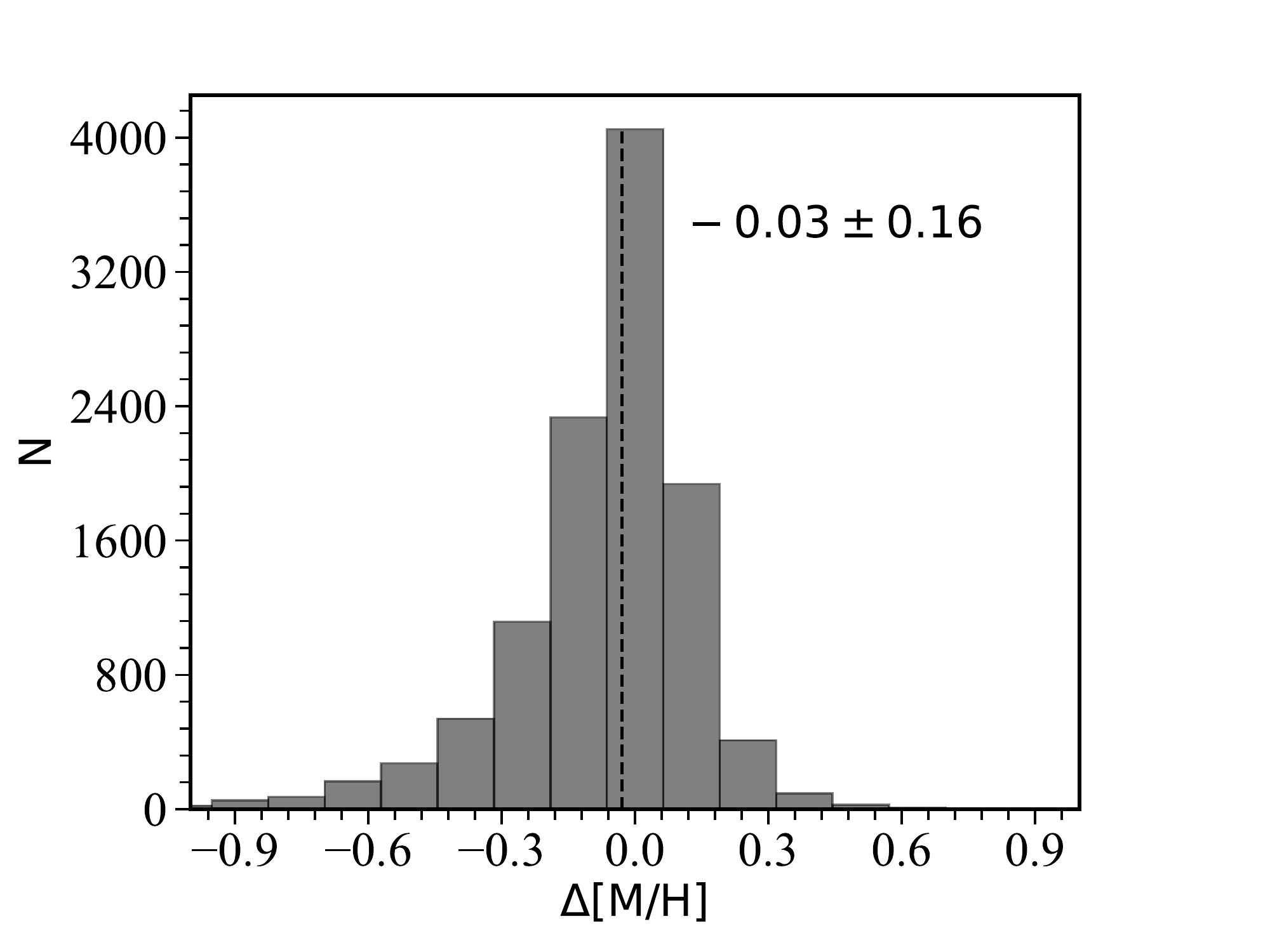}
\includegraphics[width=0.38\textwidth, trim=0.cm 0.0cm 3.0cm 1.cm, clip]{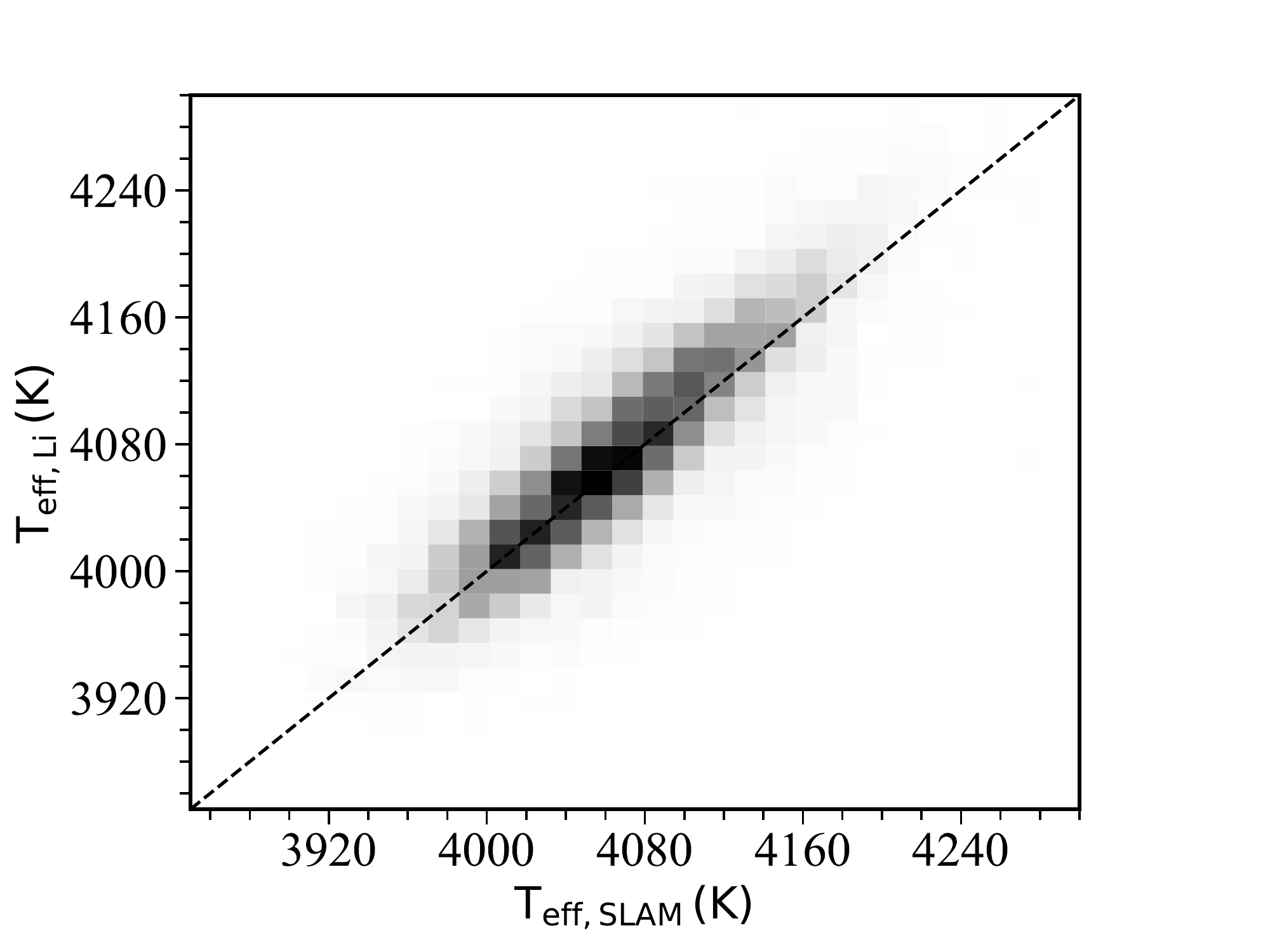}
\includegraphics[width=0.38\textwidth, trim=0.cm 0.0cm 3.0cm 1.cm,clip]{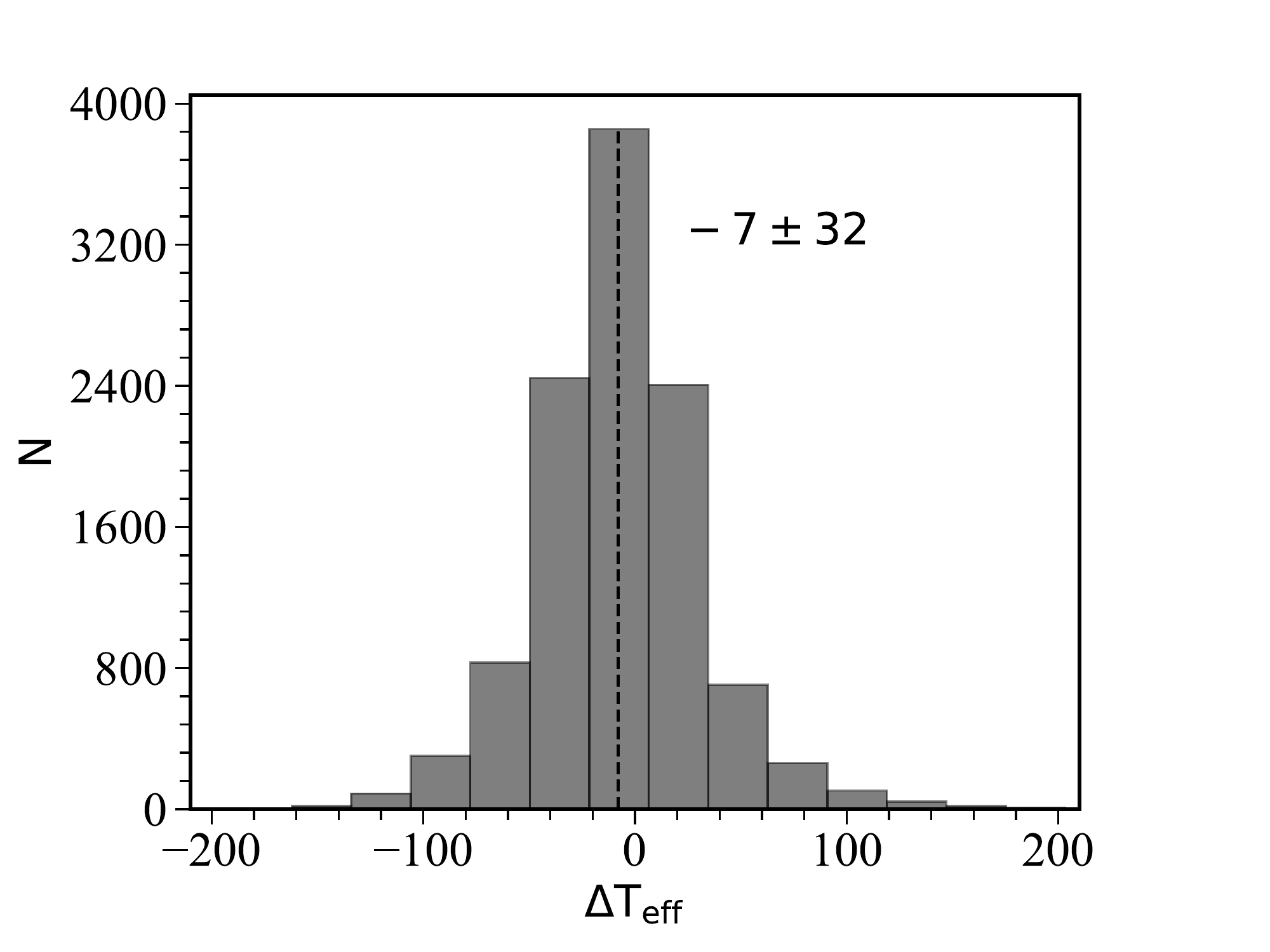}
\includegraphics[width=0.38\textwidth, trim=0.cm 0.0cm 3.0cm 1.cm, clip]{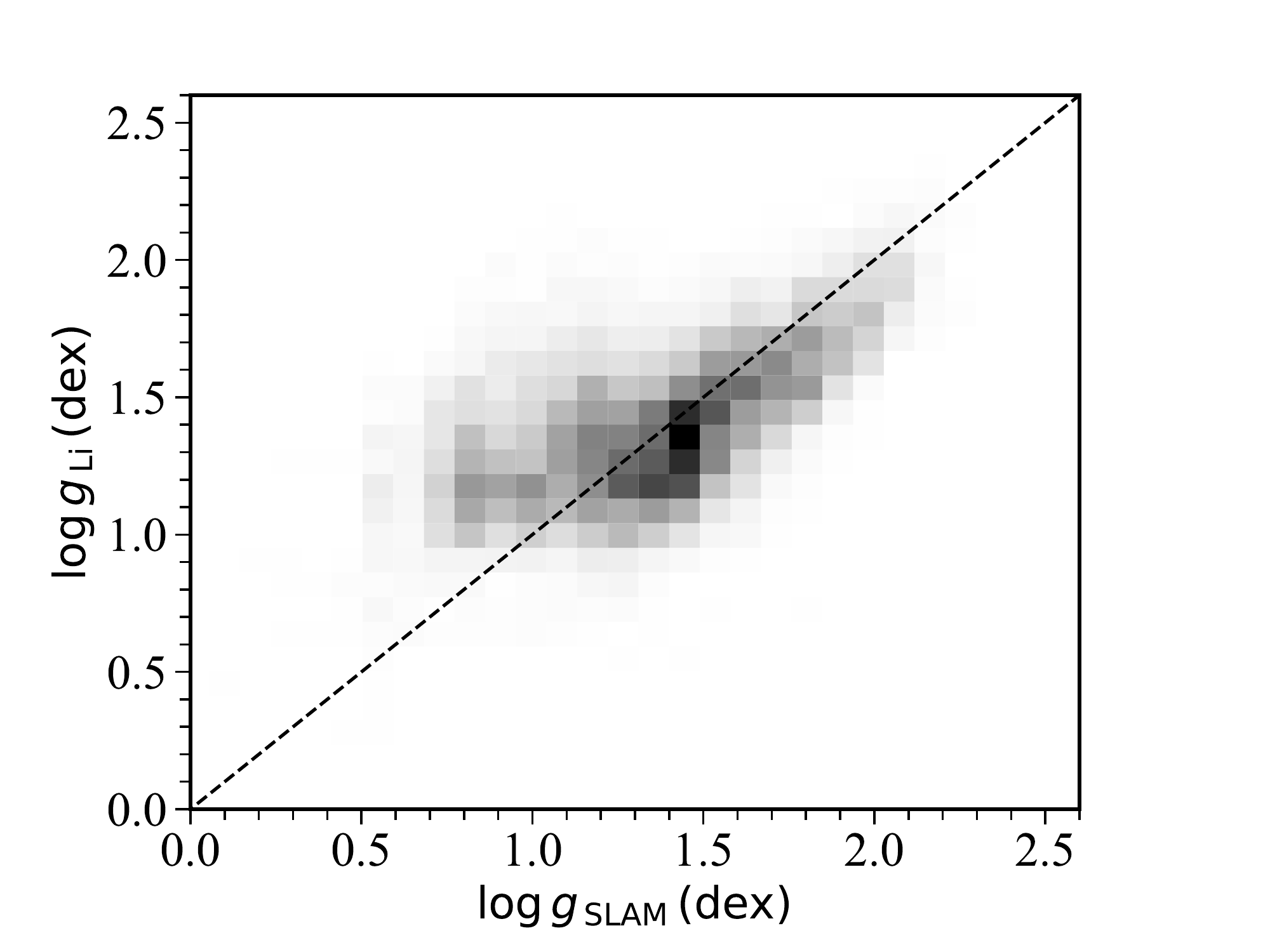}
\includegraphics[width=0.38\textwidth, trim=0.cm 0.0cm 3.0cm 1.cm,clip]{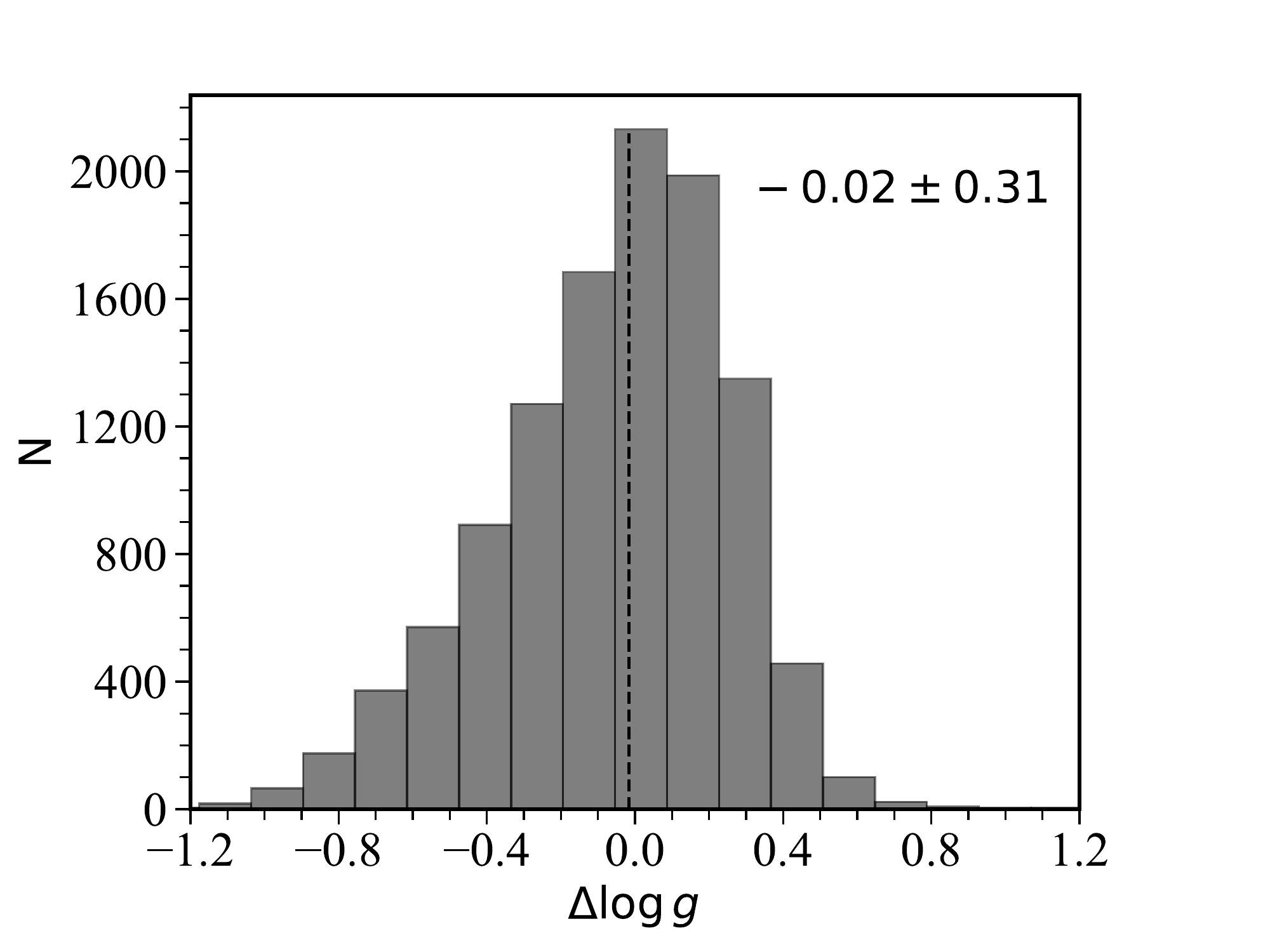}
\includegraphics[width=0.38\textwidth, trim=0.cm 0.0cm 3.0cm 1.cm, clip]{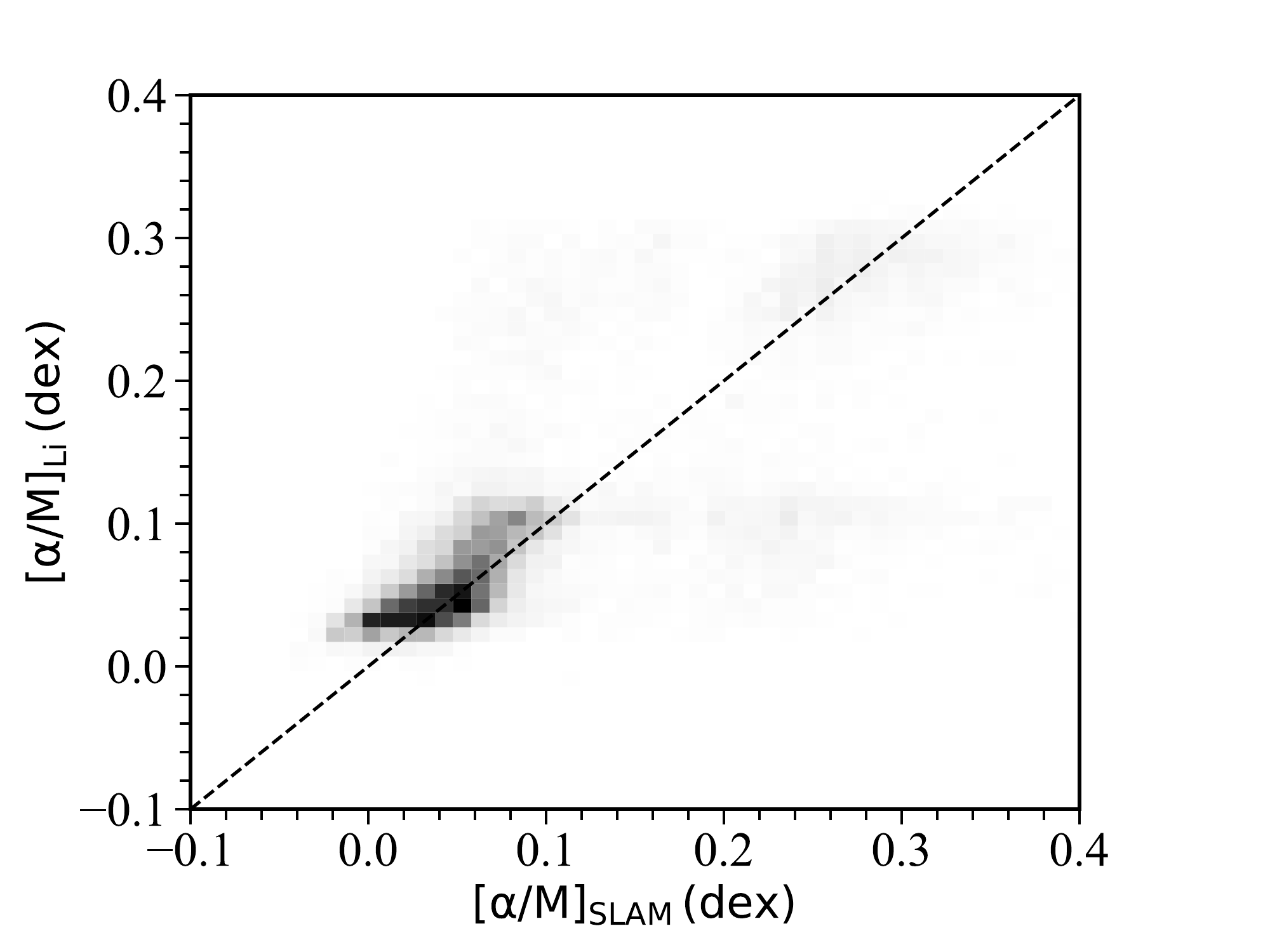}
\includegraphics[width=0.38\textwidth, trim=0.cm 0.0cm 3.0cm 1.cm,clip]{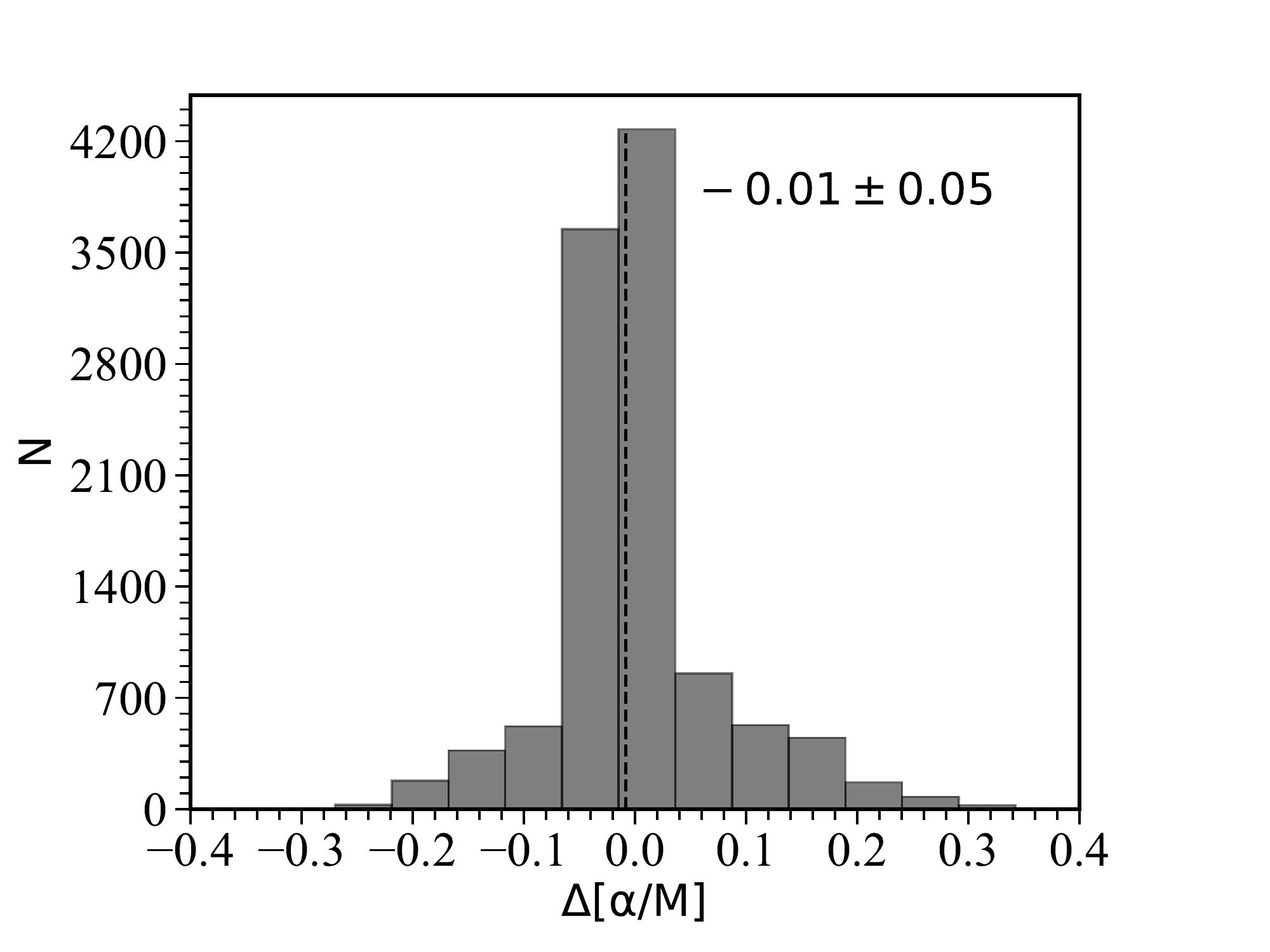}

\caption{The comparison of [M/H], $T_{\mathrm{eff}}$, log\,$g$ and 
[$\alpha$/M] of 11,132 common stars in this work and \citet{Li-2022}.
The left four panels from top to the bottom display the distributions of X$\rm_{SLAM}$ and X$\rm_{Li}$, 
where X can be [M/H], T$\rm{eff}$, log\,$g$ and [$\alpha$/M], respectively. 
The subscript 'SLAM' illustrates the parameters from this work and subscript 'Li' 
represents that from \citet{Li-2022}. The black dotted lines in the left four panels 
are the one-to-one line. The corresponding histograms of the parameter 
difference $\rm \Delta X$= X$\rm_{SLAM}$-X$\rm_{Li}$ as shown in the right four panels. 
The mean value
of $\rm \Delta X$ as marked by the 
black dotted lines in the right panels.}\label{fig:vali}
\end{figure*}

We take a comparison of four stellar parameters with \citet{Li-2022}. 
They designed a deep convolution neural network with training stellar labels 
from APOGEE DR17 to derived the $T_{\mathrm{eff}}$, log\,$g$ and other 12 chemical
abundance of 1,210,145 LAMOST DR8 giants with low resolution spectra (R\,$\sim$1800). 
11,132 common stars were obtained in our samples and \citet{Li-2022}. The comparison 
of [M/H], $T_{\mathrm{eff}}$, log\,$g$ and [$\alpha$/M] are displayed in the left 
four panels in Figure \ref{fig:vali}. It obviously shows that the parameters in this work 
are consistent with that in \citet{Li-2022}. The corresponding histograms of
$\rm \Delta X$= X$\rm_{SLAM}$-X$\rm_{Li}$ are illustrated in the right four 
panels of Figure \ref{fig:vali}. The systematic offset between $T_{\mathrm{eff,SLAM}}$ 
and $T_{\mathrm{eff,Li}}$ is 7 K with a scatter of 32 K. For [M/H], log\,$g$ and [$\alpha$/M], 
there are very small biases (0.01~0.03 dex) between this work and \citet{Li-2022} with
scatters of 0.16 dex, 0.31 dex and 0.05 dex, respectively. It demonstrates
that these four parameters in this work are consistent with that of \citet{Li-2022}.

\subsection{Distance}
In the selected M giant stars, most of them are the red giant branch
stars and a few asymptotic giant branch stars. In order to calculate the
distance, we firstly constrain the absolute magnitude for each M giant star.
Here we assume the 
absolute magnitude $K_{\mathrm{abs}}$ is 
a function $D$ of the intrinsic color index $c_0$ and metallicity $M$, i.e. $K_{\mathrm{abs}}=D(c_0, M)$. 
As showed by \cite{Majewski-2003}, \cite{Li-2016a} and Figure~\ref{fig:Train_label},  
the M giant stars are mainly tracing the metal rich components, such as the disk or the 
Sagittarius Stream, the $J$ and $K-$band magnitudes are adopted
to determine the distance to reduce the extinction effect from dust.

Assuming the $i$th star has the apparent magnitudes $J_i$ and $K_i$ and the
distance $d_i$, and suffers the extinction
$A^K_i(d_i)$, where $A^K(d_i)$ is a function of distance  
$d_i$, then we have the following relation where both 
sides are the absolute magnitude $K_{\mathrm{abs}}$,
\begin{equation}\label{Equ:AbsMag}
    K_i-A^K_i(d_i)-5\times\mathrm{log}\,(100\times d_i) = D(c_0, M)
\end{equation}
where $c_0$ is the intrinsic color index, which can be calculated by
\begin{equation}\label{Equ:ColorIndex}
c_0= (J-K)-(A^J(d)-A^K(d))
\end{equation}
Then we can find that  only the true distance $d$
satisfies the Equations (\ref{Equ:AbsMag}) and (\ref{Equ:ColorIndex}).
Here the 3D dust map from \cite{Green2019ApJ...887...93G} is adopted, which has 
a  median uncertainty around $30\%$. 

\begin{figure*}
\centering
\includegraphics[width=0.48\textwidth]{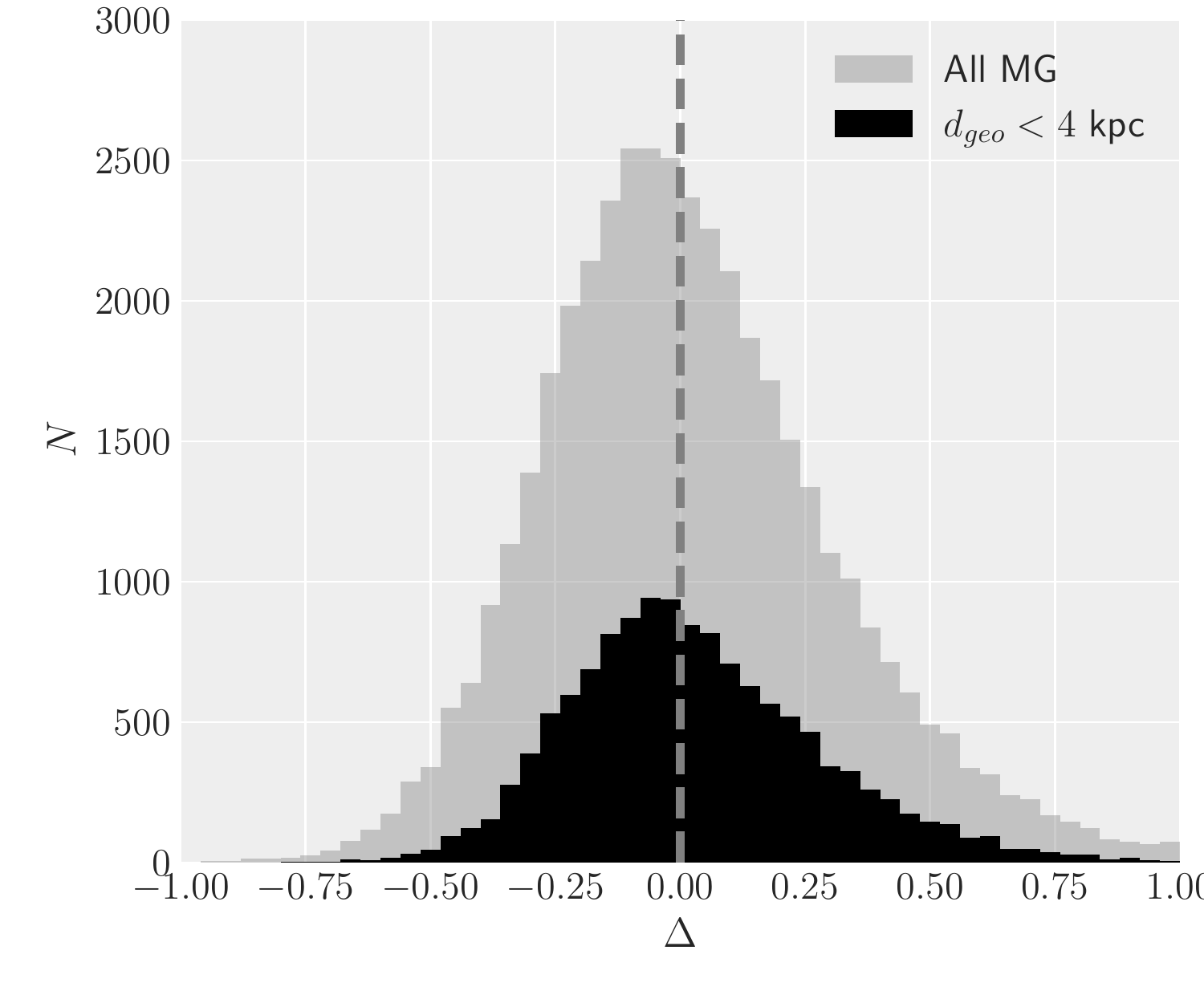}
\includegraphics[width=0.48\textwidth]{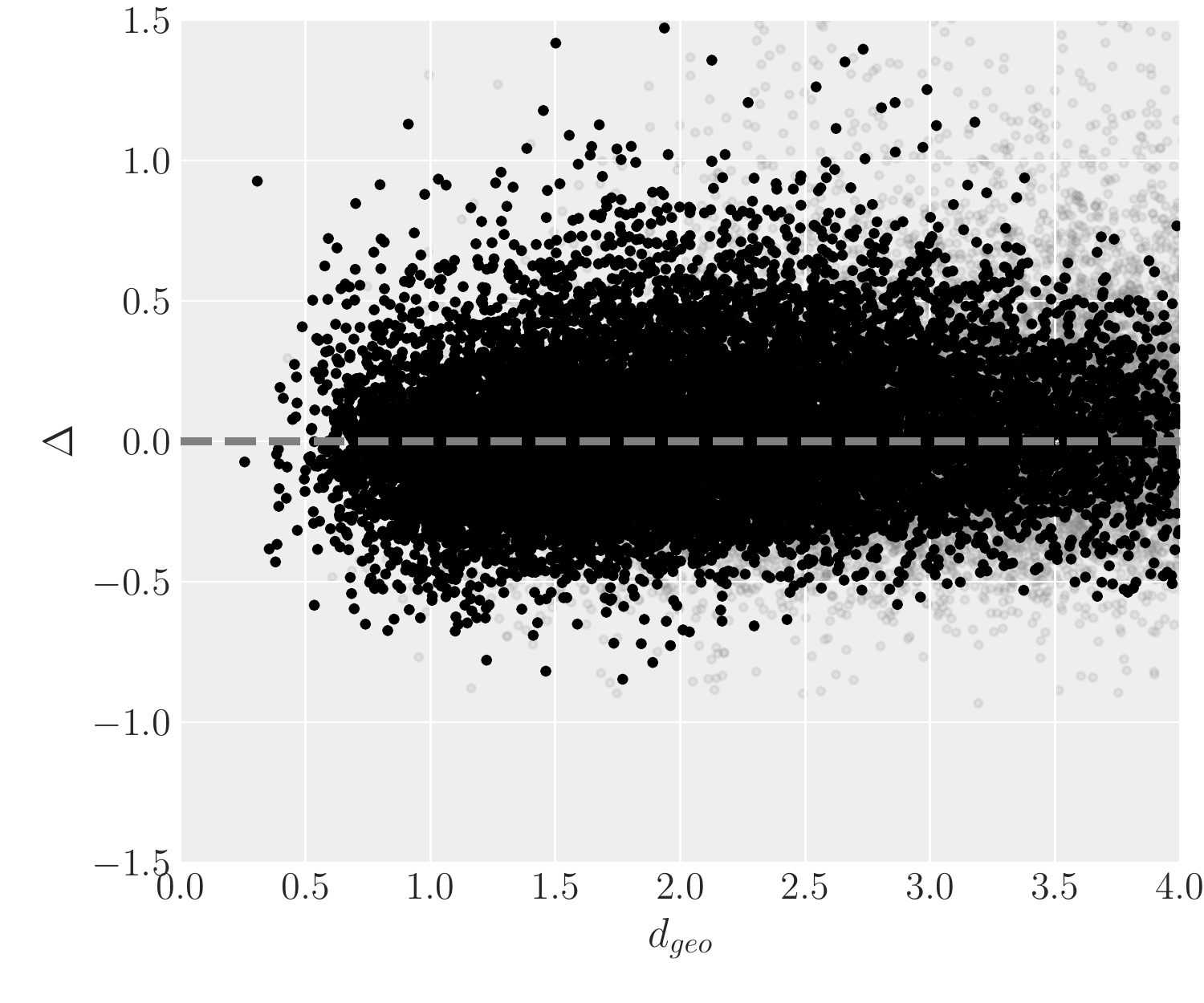}
\caption{The distance comparison with $r_{\mathrm{geo}}$ from
\citep{Bailer-Jones2021AJ....161..147B}
is showed. The histogram distribution of the relative difference
$\Delta=(d_{\mathrm{MCM}}-d_{\mathrm{geo}})/d_{\mathrm{geo}}$is showed the 
left panel. The distribution of relative difference versus the 
distance is showed in the right panel. In both panel, the gray symbols
represent the distribution of all the samples with predicted metallicity between -0.9 and 0.5.
While the black ones represent the samples with distance $d_{\mathrm{geo}}<4$ kpc and 
its signal to noise ratio larger than 20.
 }\label{fig:Dist}
\end{figure*}
Following \cite{TianH2020ApJ...899..110T}, we show the distributions of the comparison of 
the predicted distance $d_{\mathrm{MCM}}$ with 
that obtained from parallax \citep{Bailer-Jones2021AJ....161..147B}, $d_{\mathrm{geo}}$.
The subsample with reliable distances are selected with $d_{\mathrm{geo}}<4$ kpc and
the $d_{\mathrm{geo}}/\sigma_{d_{\mathrm{geo}}}>20$. Meanwhile the metallicity is also
limited to be between -0.9 and 0.5. Then the distributions of the relative distance difference
$\Delta=(d_{\mathrm{MCM}}-d_{\mathrm{geo}})/d_{\mathrm{geo}}$ of the subsample are 
showed in the Figure~\ref{fig:Dist}. The comparison shows a median difference $0.2\%$ and the 
16\% and 84\% percentage values of 21.2\% and 27.5\%., which indicate a very small system 
offset and statistical dispersion smaller than 30\%. 
In the right panel, the relative distance difference is represented versus the distance
$d_{\mathrm{geo}}$. There is not significant relation between the system offset versus
the distance.

\section{Discussion}\label{sect:App}

\begin{figure*}
\centering
\includegraphics[width=1.1\textwidth,trim=2.5cm 0.0cm 0.0cm 0.cm,clip]{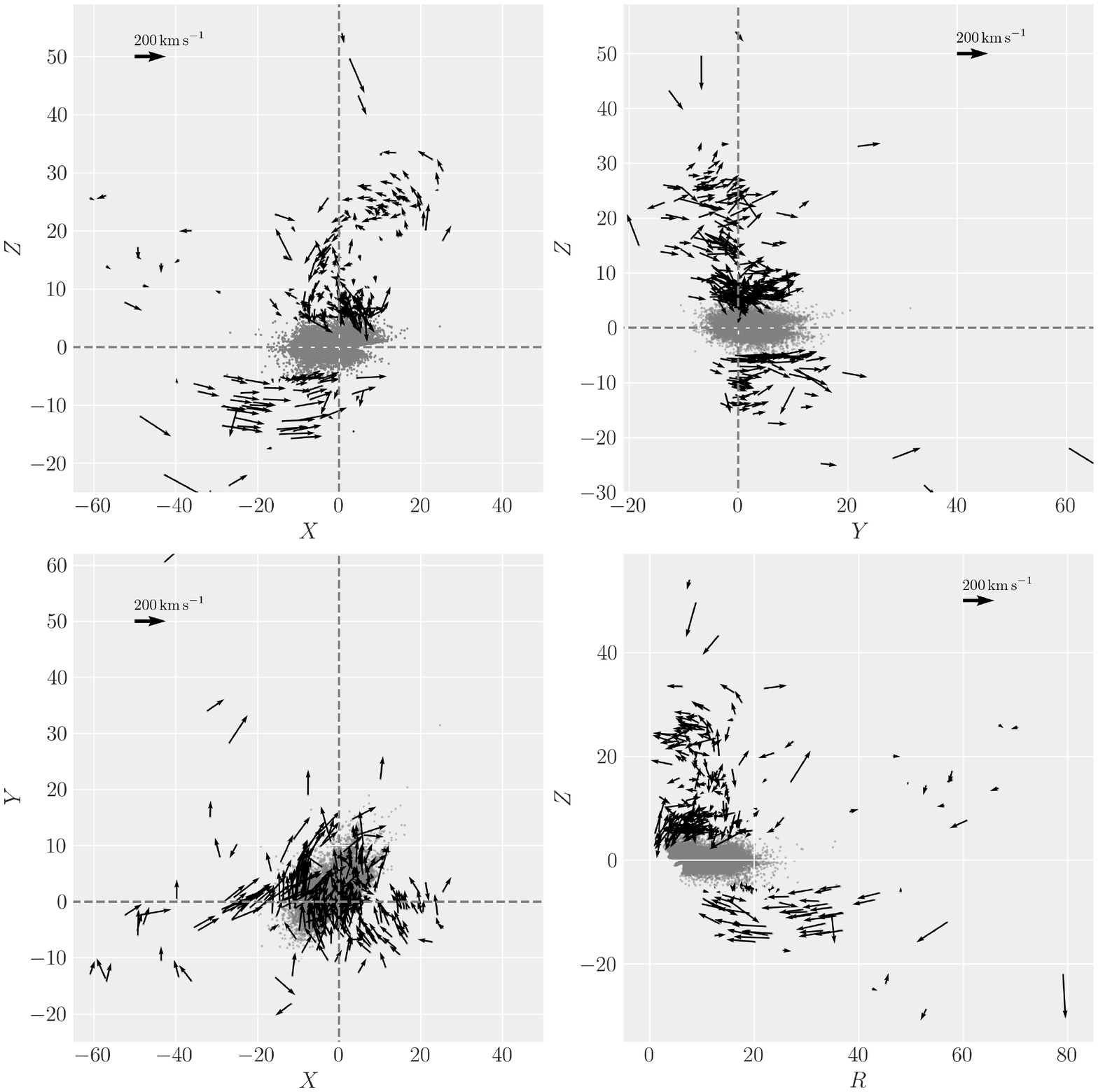}
\caption{The space distributions are showed in different spaces. 
The velocities of the samples with larger
height $|Z|>5$ kpc in the corresponding space are represented by
the arrows. The Sun is located at $(X, Y, Z)=(0,0,0)$ kpc and $(R, Z)=(8.34, 0)$ kpc.}\label{fig:POS}
\end{figure*}

With the distances and radial velocity combining the astrometric information from Gaia,
we are able to calculate the full  6D information, positions $(X, Y, Z)$ and 
velocities $(U, V, W)$. Figure~\ref{fig:POS} shows the space distributions of the M giant stars. 
The location of the Sun is represented by the  dashed lines. We can find that 
most of the M giant stars are located in the disk, $|Z|<5$ kpc, meanwhile,
there are also few M giant stars  of larger heights, which are possible the member stars of 
the Sagittarius Stream.

\subsection{Disks}

\begin{figure*}
\centering
\includegraphics[width=0.49\textwidth]{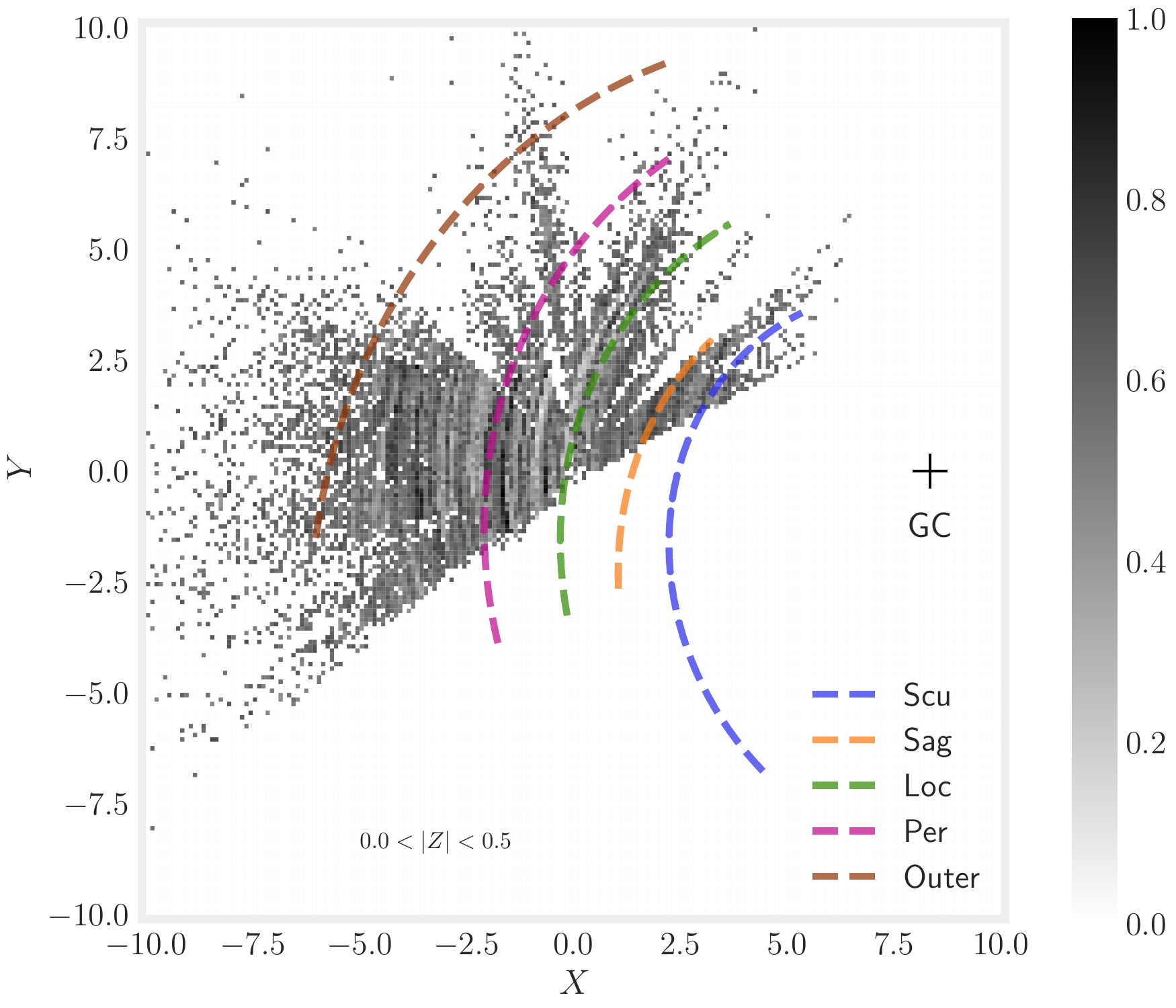}
\includegraphics[width=0.49\textwidth]{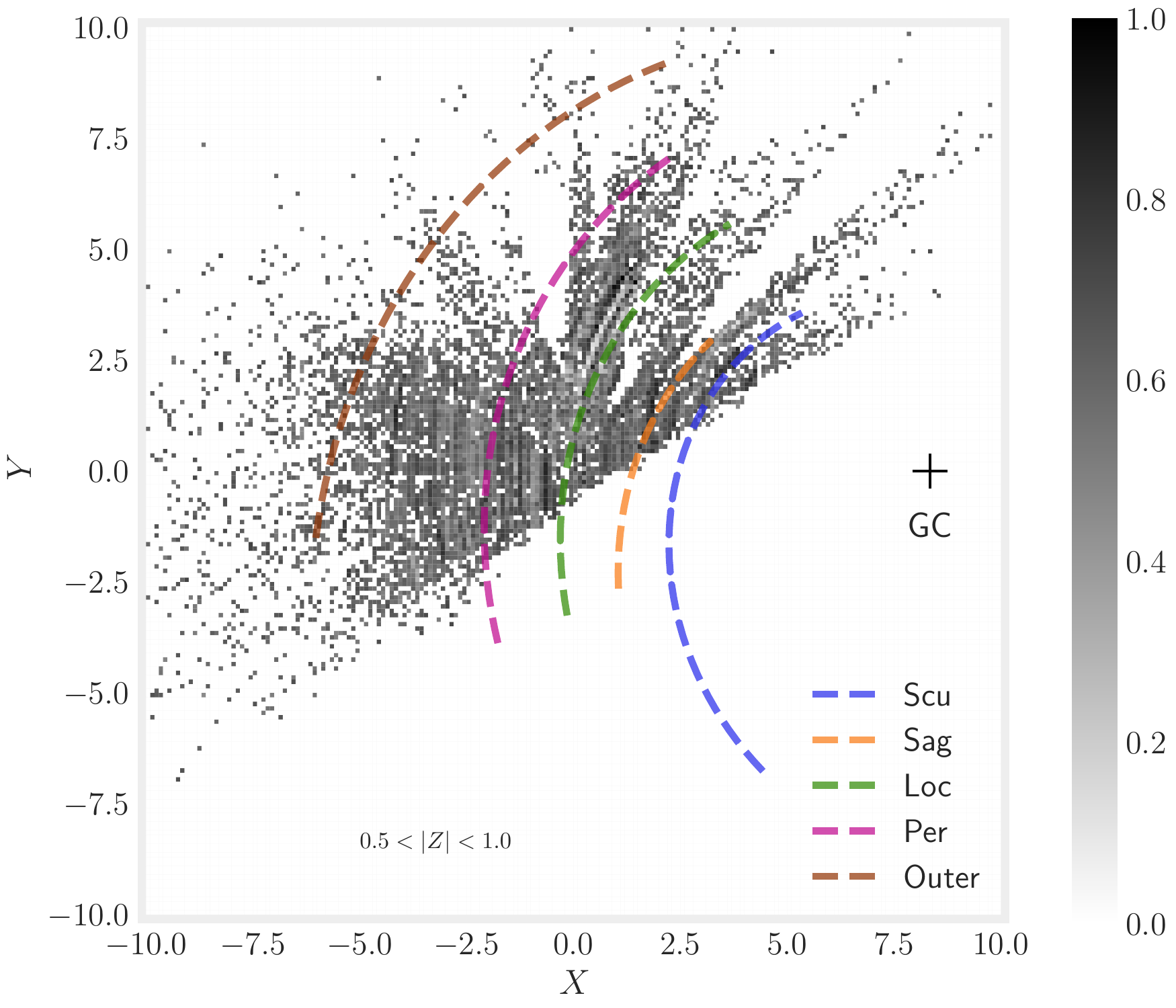}\\
\includegraphics[width=0.49\textwidth]{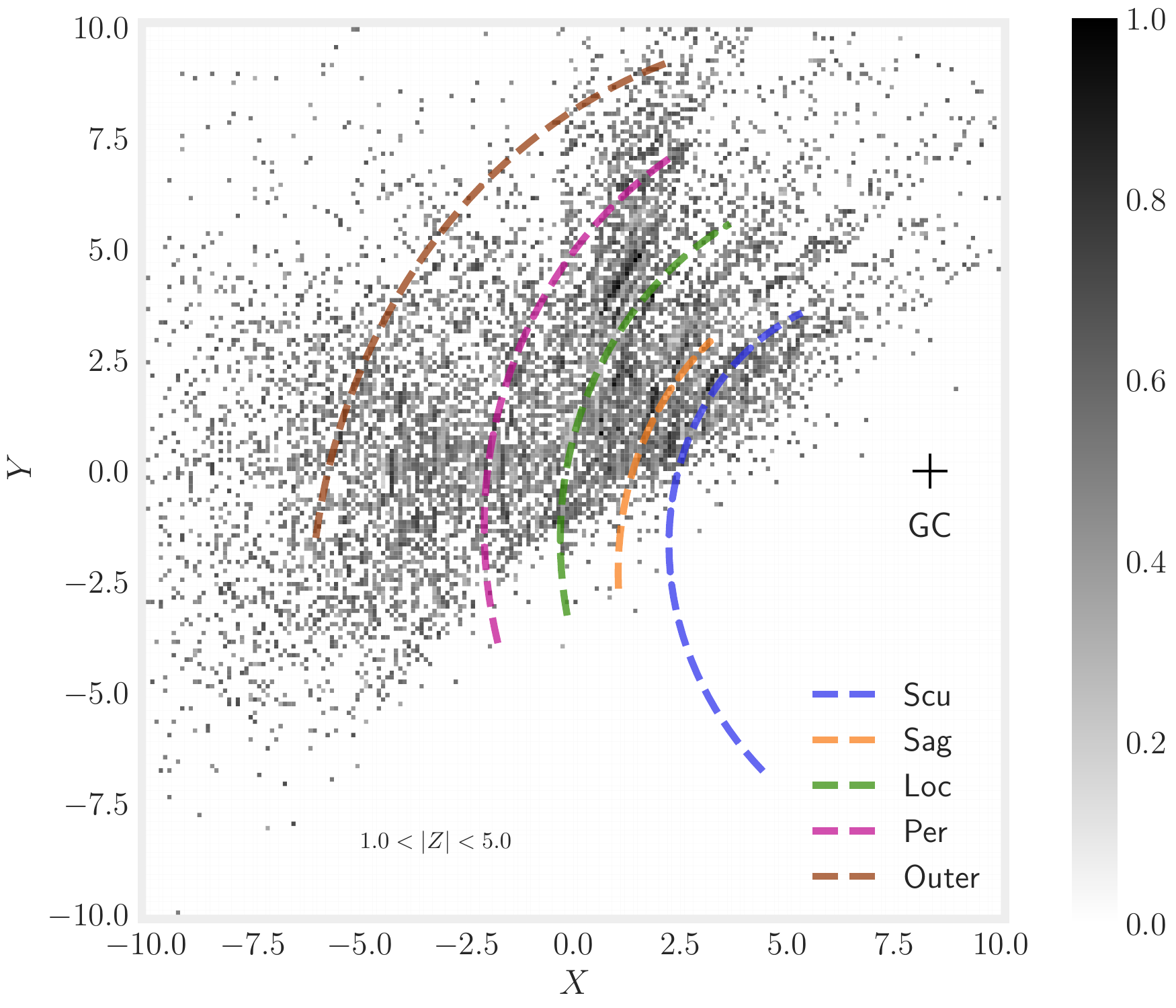}
\includegraphics[width=0.49\textwidth]{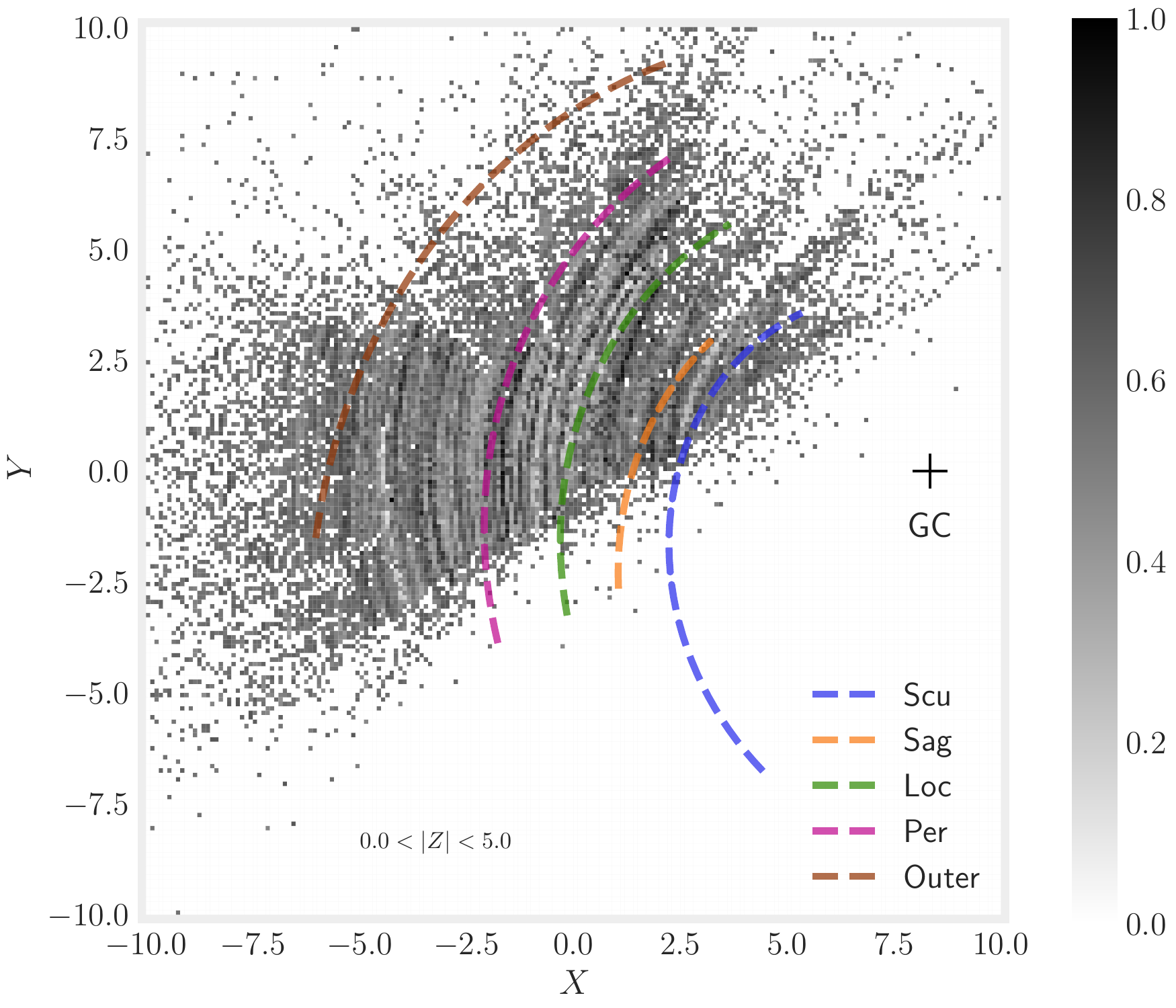}\\
\caption{Disk rotational velocity distribution after line integral convolution with 
different heights.}\label{fig:Disk}
\end{figure*}
As showed in Figure~\ref{fig:POS}, majority of our sample are the disk stars,
which are located with height smaller than $5$ kpc. Figure~\ref{fig:Disk}
shows the movements of disk traced by the M giant stars with different heights.
The five known spiral arms are represented by the dashed curves from
\cite{ChenBQ2019MNRAS.487.1400C}. The distributions of the 
line  integral convolution show clear streaming movement
which are the rotation of the disk,
especially the thin disk represented by the 
subsample with $|Z|<0.5$ kpc. 

\begin{figure*}
\centering
\includegraphics[width=0.49\textwidth]{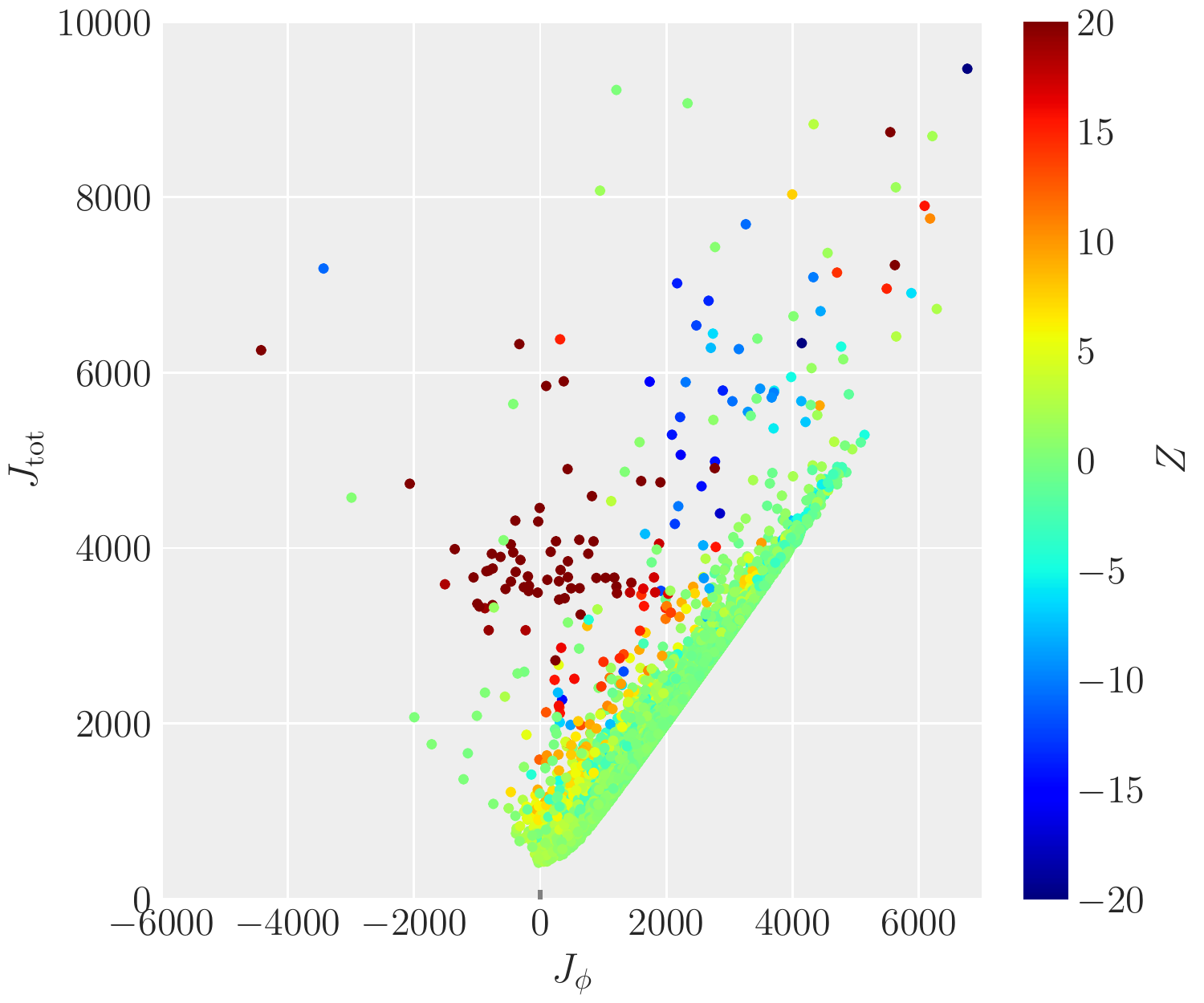}
\includegraphics[width=0.49\textwidth]{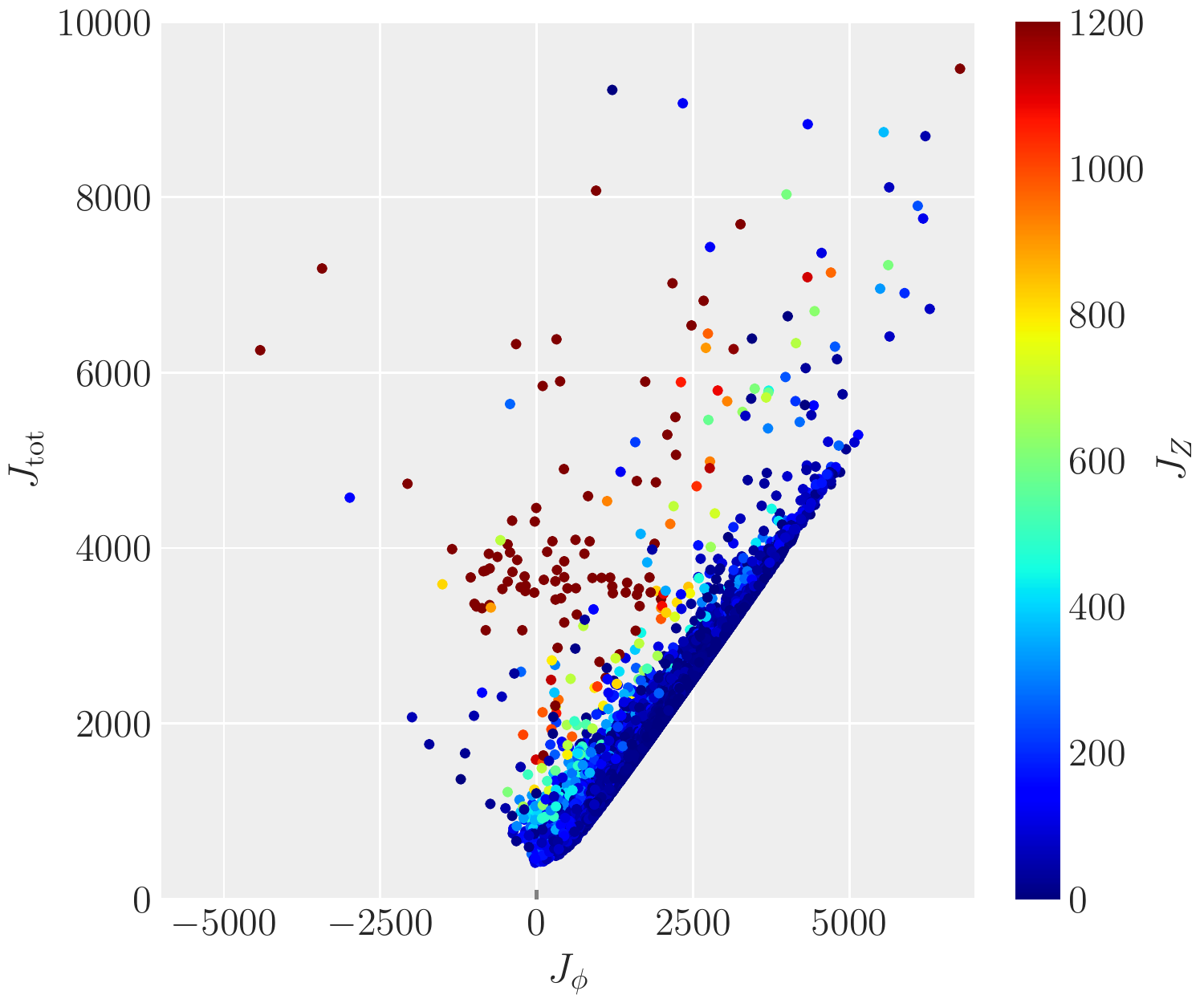}
\caption{The distribution of the M giant stars in the 
action space $J_{\phi}$ versus $J_{\mathrm{tot}}$ are showed,
which are color-coded by the  height $Z$ and action $J_Z$ in
the left and right panels, respectively.}\label{fig:AJ}
\end{figure*}
Figure~\ref{fig:AJ} shows the phase space distribution of the 
M giant stars,  $J_{\phi}$ versus $J_{\mathrm{tot}}$. The samples
are color-coded by the height $Z$ to the disk plane and the action 
$J_Z$ in the left and right panels, respectively. The action distributions
also prove that the majority of the M giant stars are concentrated in
the thin cyan and blue belt in the left and right panels, respectively, 
which are the disk with low height $|Z|$ and action $J_Z$.

\subsection{Sagittarius Stream}
\begin{figure*}
\centering
\includegraphics[width=1.\textwidth,trim=2.cm 5.cm 2.0cm 1.cm, clip]{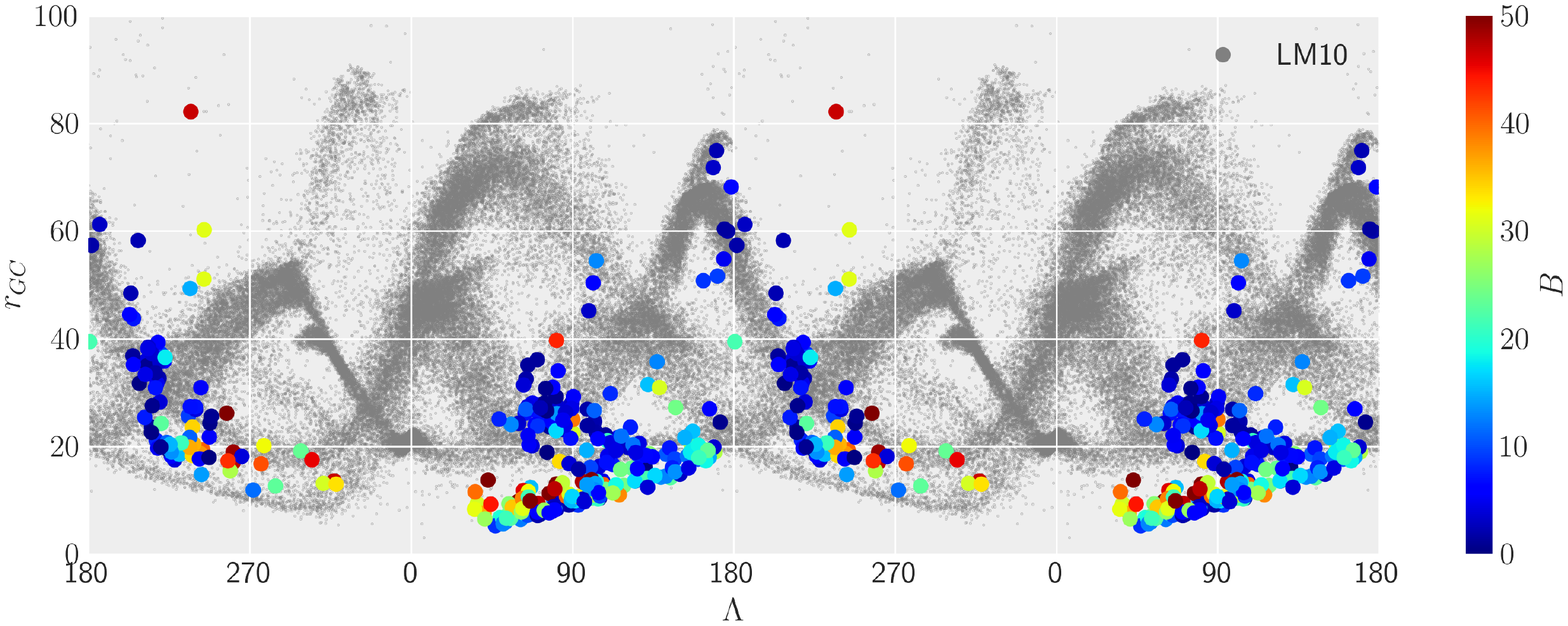}
\caption{The distance distribution versus longitude $\Lambda$ of the M 
giant stars with height larger than $5$ kpc is showed which 
are color coded by the latitude $B$. The samples of the simulated Sagittarius
Stream from LM10 are showed with gray 
symbols. }\label{fig:Sag}
\end{figure*}

Another application for the M giant sample 
is to trace the Sagittarius Stream \citep{Majewski-2003,Li-2016a}.
From the Figures~\ref{fig:POS} and \ref{fig:AJ}, there are also a few stars with 
larger heights $|Z|>5$ kpc, whose movements are also represented by the arrows.
From previous studies
\citep{Majewski-2003, Li-2016a}, the Sagittarius Stream has 
a significant contribution to the M giant stars with high latitude.
In Figure~\ref{fig:POS}, those M giant stars with larger heights
show significant bulk motions, especially those stars with $Z$ around $25$ kpc and $-10$ kpc. 
That is more clear in
Figure~\ref{fig:AJ}, where those samples are of larger total actions
$J_{\mathrm{tot}}$ and small angular momentum $J_\phi$, which are
completely different with those belong to the
disk. To further illustration, we convert the coordinate
$(\alpha, \delta)$ to that based on the Sagittarius Stream plane
$(\Lambda, B)$, where the north pole $B=90^\circ$ is set to the same direction of 
the normal of the orbit plane of the Sagittarius Stream. 
Then Figure~\ref{fig:Sag} shows the distance variance 
versus the longitude $\Lambda$ color-coded by the latitude $|B|$.
The mock data for the Sagittarius Stream from \cite{Law2010ApJ...714..229L} 
is also showed
with gray dots. We find that the distance distribution of those
M giant stars fits the model well. There are also few nearby stars
offset the model which are possible the flared disk stars with 
$|Z|>5$ kpc. That can be proved by the action distributions as
showed in Figure~\ref{fig:AJ}.

\section{Summary}\label{sect:conclu}
With similar method with \cite{Zhong2015RAA....15.1154Z}, \cite{Zhong2019ApJS..244....8Z},
Li et al (in preparation) select M giant stars from LAMOST DR9 with a very high purity.
Many of those low temperature M giant stars are not given the stellar parameters and
the radial velocities in the official catalog.
In this work, we revise the spectra of those M giant stars and constrain
the stellar parameter 
$(T_{\mathrm{eff}}$,$\mathrm{log}\,g$, $\mathrm{[M/H]})$, the 
chemical abundance  $\mathrm{[\alpha/M]}$
and the radial velocities with uncertainties of
$(57\, \mathrm{K}, 0.25\, \mathrm{dex}, 0.16\, \mathrm{dex}, 0.06\, 
\mathrm{dex}) $ 
and $4.6$ km s$^{-1}$, respectively. With those information,
we are able to calculate the full 6D information of M giants, and further to study the Milky Way disks
and the Sagittarius Stream. 
Combining the geometric and phase space distributions, the disk and the Sagittarius Stream can 
be well separated. This value added sample will provide a pure sample for  the chemical and kinematic
 studies for the disk and the Sagittarius Stream.

\bibliographystyle{aasjournal}
\bibliography{bibtex}

\section{acknowledge}

H.T. is supported by Beijing Natural Science Foundation with grant No. 1214028 and 
the National Natural Science Foundation of China (NSFC)  under grants 12103062.
J.L. would like to acknowledge the NSFC under grant 12273027 and China West Normal University grants 17YC507.
C.L. thanks the National Natural Science Foundation of China (NSFC) with grant Nos.11835057 and 
the National Key R\&D Program of China No. 2019YFA0405501.
J.R.S. is supported by the National Key R\&D Program of China No. 2019YFA0405502
and the National Natural Science Foundation of China under grant Nos. 12090040, 12090044, 11833006.
M.Y. gratefully acknowledges support from the National Natural Science Foundation of China (Grant No.12133002).
Guoshoujing Telescope (the Large Sky Area Multi-Object Fiber Spectroscopic Telescope LAMOST) is a 
National Major Scientific Project built by the Chinese Academy of Sciences. Funding for the project has 
been provided by the National Development and Reform Commission. LAMOST is operated and 
managed by the National Astronomical Observatories, Chinese Academy of Sciences.

\appendix  \label{app:A}
\section{Catalog sample}
Table~\ref{col:all} lists the information in our catalog, including the observational ID, 
(obsid), the mean signal to noise ratio (SNR) of LAMOST spectra and the coordinates (ra\_obs, dec\_obs) from LAMOST. The stellar parameters, the distance,
the radial velocities and their uncertainties are also provided. The whole catalog 
is available to download through the link \url{https://nadc.china-vo.org/res/r101196/}.
\begin{table*} 
    \centering
    \caption{Catalog description of $\sim$ 43,000 M giants}  \label{col:all}
\begin{tabular}{p{2.0cm}|p{1.5cm}|p{8.0cm}}
\hline\hline
Column & units & Description \\ \hline
{obsid}                          &         & LAMOST observe id   \\
{ra\_obs}                                  & deg     & right ascension from LAMOST  \\
{dec\_obs}                                 & deg     & declination from LAMOST \\
{SNR}                    &             &  mean signal to noise ratio of LAMOST spectra \\
{[M/H]}               & dex     & [M/H] from SLAM \\
{[M/H]$\rm_{err}$}               & dex     & [M/H] uncertainty\\
{$T_{\mathrm{eff}}$}               & K     & T$\rm_{eff}$ from SLAM \\
{${T_{\mathrm{eff}}}\rm_{err}$}               & K     & $T_{\mathrm{eff}}$ uncertainty\\
{$\mathrm{log}\,g$}               & dex     & $\mathrm{log}\,g$ from SLAM \\
{$\mathrm{log}\,g\rm_{err}$}               & dex     & $\mathrm{log}\,g$ uncertainty\\
{$\mathrm{[\alpha/M]}$}               & dex     & $\mathrm{[\alpha/M]}$ from SLAM \\
{$\mathrm{[\alpha/M]}\rm_{err}$}               & dex     & $\mathrm{[\alpha/M]}$ uncertainty\\
{rv}                    &    $\rm km\,s^{-1}$         &  radial velocity  \\
{Distance}                    &    kpc         &  Distance of M giants  \\

\hline
\hline
\end{tabular}
\end{table*}

\label{lastpage}

\end{document}